\documentclass[modern]{aastex631}
\usepackage{amsmath}
\usepackage{microtype}
\usepackage{soul}

\newcommand{\mgfe}[0]{[{\rm Mg/Fe}]}

\newcommand{\femg}{[{\rm Fe}/{\rm Mg}]} 
\newcommand{\xmg}{[{\rm X}/{\rm Mg}]} 
\newcommand{\mgh}{[{\rm Mg}/{\rm H}]}
\newcommand{\feh}[0]{[{\rm Fe/H}]}

\newcommand{\xh}{[{\rm X}/{\rm H}]}
\newcommand{\logg}{\log(g)}
\newcommand{\teff}{T_{\rm eff}}

\newcommand{\qcc}{q_{{\rm CC,}j}^{Z}}
\newcommand{\qccFe}{q_{{\rm CC, Fe}}^{Z}}
\newcommand{\dqccFe}{dq_{{\rm CC, Fe}}^{Z}/dZ}
\newcommand{\qIa}{q_{{\rm Ia,}j}^{Z}}
\newcommand{\qIaFe}{q_{{\rm Ia,Fe}}^{Z}}
\newcommand{\Acc}{A^{\rm CC}_{i}}
\newcommand{\AIa}{A^{\rm Ia}_{i}}
\newcommand{\fcc}{f_{ij}^{\rm CC}}

\newcommand{\qccX}[1]{q_{{\rm CC, #1}}^{Z}}
\newcommand{\qIaX}[1]{q_{{\rm Ia, #1}}^{Z}}
\newcommand{\fccX}[1]{f_{i, {\rm #1}}^{\rm CC}}

\submitjournal{ApJ}

\newcommand{\name}{\textsl{KPM}} 
\newcommand{\documentname}{\textsl{Article}}

\shorttitle{a data-driven model for nucleosynthesis}
\shortauthors{Griffith \& Hogg}
\renewcommand{\twocolumngrid}{} 
\renewcommand{\paragraph}[1]{\bigskip\par\noindent{\textbf{#1}}~---}
\sloppypar

\graphicspath{{./}{Figures/}}
\begin{document}

\title{\name: A flexible and data-driven K-process model for nucleosynthesis}

\correspondingauthor{Emily J. Griffith}
\email{Emily.Griffith-1@colorado.edu}

\author[0000-0001-9345-9977]{Emily J. Griffith}
\altaffiliation{NSF Astronomy and Astrophysics Postdoctoral Fellow}
\affiliation{Center for Astrophysics and Space Astronomy, Department of Astrophysical and Planetary Sciences, University of Colorado, 389~UCB, Boulder,~CO 80309-0389, USA}

\author[0000-0003-2866-9403]{David W. Hogg}
\affiliation{Center for Cosmology and Particle Physics, Department of Physics, New York University, 726~Broadway, New~York,~NY 10003, USA}
\affiliation{Max-Planck-Institut f{\"u}r Astronomie, K{\"o}nigstuhl 17, D-69117 Heidelberg, Germany}
\affiliation{Center for Computational Astrophysics, Flatiron Institute, 162 Fifth Avenue, New York, NY 10010, USA}

\author{Julianne J. Dalcanton}
\affiliation{Center for Computational Astrophysics, Flatiron Institute, 162 Fifth Avenue, New York, NY 10010, USA}
\affiliation{Department of Astronomy, Box 351580, University of Washington, Seattle, WA 98195}

\author[0000-0001-5388-0994]{Sten Hasselquist}
\affiliation{Space Telescope Science Institute, 3700 San Martin Drive, Baltimore, MD 21218}

\author[0000-0003-1124-7378]{Bridget Ratcliffe}
\affiliation{Leibniz-Institut f\"{u}r Astrophysik Potsdam (AIP), An der Sternwarte 16, 
14482 Potsdam, Germany}

\author[0000-0001-5082-6693]{Melissa Ness}
\affiliation{Center for Computational Astrophysics, Flatiron Institute, 162 Fifth Avenue, New York, NY 10010, USA}
\affiliation{Department of Astronomy, Columbia University, Pupin Physics Laboratories, New York, NY 10027, USA}

\author[0000-0001-7775-7261]{David H. Weinberg}
\affiliation{The Department of Astronomy and Center of Cosmology and AstroParticle Physics, The Ohio State University, Columbus, OH 43210, USA}

\begin{abstract}\noindent
The element abundance pattern found in Milky Way disk stars is close to two-dimensional, dominated by production from one prompt process and one delayed process. 
This simplicity is remarkable, since the elements are produced by a multitude of nucleosynthesis mechanisms operating in stars with a wide range of progenitor masses. 
We fit the abundances of 14 elements for 48,659 red-giant stars from APOGEE DR17 using a flexible, data-driven K-process model---dubbed \name. 
In our fiducial model, with $K=2$, each abundance in each star is described as the sum of a prompt and a delayed process contribution. 
We find that \name with $K=2$ is able to explain the abundances well, recover the observed abundance bimodality, and detect the bimodality over a greater range in metallicity than previously has been possible. 
We compare to prior work by \citet{weinberg2022}, finding that \name produces similar results, but that \name better predicts stellar abundances, especially for elements C+N and Mn and for stars at super-solar metallicities. 
The model fixes the relative contribution of the prompt and delayed process to two elements to break degeneracies and improve interpretability; we find that some of the nucleosynthetic implications are dependent upon these detailed choices. 
We find that moving to four processes adds flexibility and improves the model's ability to predict the stellar abundances, but doesn't qualitatively change the story. 
The results of \name will help us to interpret and constrain the formation of the Galaxy disk, the relationship between abundances and ages, and the physics of nucleosynthesis.
\end{abstract}

 
\section{Introduction}\label{sec:intro}

After hydrogen, helium, lithium, and beryllium, all other naturally occurring elements are made in stars, supernovae, and the collisions of stars.
Stellar surface abundances---the abundances measured by taking a spectrum of a stellar photosphere---are thought to deliver a relatively unprocessed record of the element abundances in the gas from which the star formed \citep[though see, e.g.,][]{pinsonneault2001, oh2018, souto2019, vincenzo2021b}.
These birth abundances were set by a combination of nucleosynthetic processes involved in making heavy atomic nuclei, and astrophysical processes involved in delivering atoms from stellar interiors to star-formation sites \citep[e.g.,][]{johnsonja2020}.
Thus nuclear physics and a wide swath of astrophysics are critically intertwined in our understanding of stellar surface abundances, motivating theoretical, experimental, and observational work.

At the present day, stellar surface abundances are not very well explained by purely \textsl{ab initio}, physics-driven models.
Theoretical yields vary from data set to data set, as they are dependent on progenitor properties and explosion assumptions \citep[e.g.,][]{rybizki2017, blancato2019, buck2021, griffith2021b}. 
The wide parameter space of progenitor and supernova models coupled with uncertainties in reaction rates and explosion physics hinder the creation of an accurate nucleosynthetic model from theory alone.
In the long run, it is incumbent upon us to understand these issues and correct the assumptions or calculations underlying our nucleosynthetic and astrophysical models.
In the short run, however, we gather data---tens of millions of abundance measurements on millions of stars in different astronomical surveys such as RAVE \citep{steinmetz2006}, SEGUE \citep{yanny2009}, LAMOST \citep{luo2015}, Gaia-ESO \citep{gilmore2012, gilmore2022}, APOGEE/MWM \citep{majewski2017}, GALAH \citep{desilva2015}, and H3 \citep{conroy2019}.
This raises the question: \emph{Can we take a data-driven approach to nucleosynthesis?}

In this \documentname{}, we build a purely data-driven model for the surface element abundances observed in stars.
We treat each star as being a linear combination of nucleosynthetic processes, beginning with one that is primarily responsible for the $\alpha$-element Mg \citep[prompt enrichment, such as core-collapse supernovae or CCSN, e.g.,][]{andrews2017}, and one that is primarily \emph{not} responsible for Mg (delayed enrichment, such as Type-Ia supernovae or SNIa).
Beyond these up-front assumptions, we try to be agnostic about how the elements are produced.

We build upon the work of \citet[][hereafter G22]{griffith2019, griffith2022} and \citet[][hereafter W19, W22]{weinberg2019, weinberg2022}, who used the bimodality in [Mg/Fe] vs. [Fe/H]\footnote{Where $[{\rm X}/{\rm Y}] = \log({\rm X}/{\rm Y}) - \log({\rm X}/{\rm Y})_{\odot}$ and $({\rm X}/{\rm Y})_{\odot}$ is the solar abundance ratio.} \citep[e.g.,][]{fuhrmann1998, bensby2003, adibekyan2012} to separate stars into populations with high and low SNIa enrichment. As established in \citet{griffith2019}, these populations are referred to as high-Ia and low-Ia, to reflect their enrichment origins, instead of the traditional low-$\alpha$ and high-$\alpha$ nomenclature. We adopt this updated naming convention in this \documentname. Using the median [X/Mg] vs. [Mg/H] abundance trends, these prior works explain data from the GALAH\footnote{GALAH = GALactic Archaeology with HERMES.} and SDSS-IV APOGEE\footnote{APOGEE = Apache Point Observatory Galactic Evolution Experiment, part of the Sloan Digital Sky Survey} surveys, respectively, with a two-process model. Because the median abundance trends in [X/Mg] vs. [Mg/H] space are largely insensitive to aspects of chemical evolution, such as outflows and variations in star formation history (W19), the population abundance trends are set by the nucleosynthetic processes and can be used to empirically constrain Galactic enrichment. 

These works, as well as \citet{ting2022} and \citet{ratcliffe2023}, find that the Milky Way stellar abundances are well fit by two components, grounded in [Fe/H] and [Mg/Fe], down to residuals of 0.01 to 0.03 dex for the most precisely measured elements and 0.05 to 0.1 dex for elements (such as Na, C, and Ce) with large measurement errors. Simultaneously, \citet{frankel2018} and \citet{ness2022} have found that disk abundances are also well described by a two-component model of birth radius and age. Correlations between two-process model parameters and stellar ages and kinematics (W22) as well as the success of a two-component model of [Fe/H] and age in predicting APOGEE abundances \citep{ness2019} suggest that these two 2-dimensional models are somehow interconnected. 

Beyond standard CCSN and SNIa enrichment, many elements have contributions from additional nucleosynthetic processes, such as the rapid ($r$) and slow ($s$) neutron capture processes \citep[e.g.,][]{arlandini1999, bisterzo2014} in asymptotic giant branch (AGB) stars \citep[e.g.,][]{simmerer2004, karakas2016}, merging neutron stars \citep[e.g.,][]{kilpatrick2017}, or atypical supernova explosions \citep[e.g.,][]{nomoto2013}. After predicting stellar abundances from $\feh$ and $\mgfe$, \citet{ting2022} identify correlated abundance residuals that are unexplained by observational uncertainties, indicative of additional nucleosynthetic processes that standard disk CCSN and SNIa enrichment cannot explain. Results from G22 and W22 support this conclusion, and both works attempt to add additional processes to their models to account for non-CCSN and non-SNIa enrichment, though in a restrictive manner. Other sources of abundance scatter, such as stochastic sampling of the Initial Mass Function (IMF), IMF variations, and bursty star formation history could also cause deviations away from a two-process model \citep{belokurov2018, griffith2023}.

To date, survey abundances have not been fully exploited to create a data-driven model of nucleosynthesis. While works such as \citet{ting2012}, \citet{casey2019}, and \citet{ratcliffe2020} effectively use clustering algorithms to identify elements with like sources and reduce abundance dimensionality, the results are difficult to translate into a model of nucleosynthesis. Clustering components can be linked to nucleosynthesis sources and enrichment history, but have not yet been used to describe the enrichment of a single star.

In this work, our main innovations are to relax the assumptions made in G22 and W22, to be more agnostic about the nucleosynthetic processes, and to be more principled with the measurements or inferences from data.
In the $K$ Process Model (\name{}), we find the intersection between reliable facts about nucleosynthesis and good abundance measurements to build an edifice of Galactic enrichment.
The model is hierarchical, in that it learns some parameters (process vectors) that are shared across all stars, but different for each element, and some parameters (process amplitudes) that are shared across all elements, but different for each star.
The parameters output by our model can thus be used as de-noised abundance labels; these will sharpen relationships between abundances and stellar parameters (including birth location and time). Our main contribution is to construct a data-driven model for nucleosynthesis that has good statistical properties, only enforcing a small number of constraints to break the degeneracies that arise in models of this form. All other \name{} parameters are set by the data with no fixed normalization.

This paper is organized as follows. In Section~\ref{sec:model} we present the assumptions and the implementation of \name{}. In Section~\ref{sec:data} we describe the APOGEE data sample employed in this paper.  We apply \name{} to the APOGEE data in Section~\ref{sec:fiducial} and compare our results to those of W22 in Section~\ref{subsec:w22}. In Section~\ref{sec:variations} we explore variations from the fiducial model, changing our assumptions about Fe production as well as the number of model components. Finally, we discuss and summarize our results in Section~\ref{sec:discussion}.

\section{The $K$-Process Model}\label{sec:model}

As in W22 and G22, we propose that all stellar abundances can be generated by a combination of $K$ nucleosynthetic processes.
In this picture, each element has $K$ metallicity-dependent process vector components that are shared across the full stellar sample, while each star individually has $K$ process amplitudes, which apply across all elements, such that the expected logarithmic abundance of element $j$ relative to H in star $i$ ($m_{ij}$) is defined as:
\begin{equation}\label{eq:mij_k}
    m_{ij} = \log_{10} \, \sum^K_{k=1} A_i^k \, q_{k,j}^Z ~.
\end{equation}
Each star $i$ has $K$ process amplitudes ($A^k_i$)
and each element $j$ has $K$ metallicity dependent process vector components ($q_{k,j}^{Z}$). 
The $Z$ superscript denotes the dependence of the process vectors on metallicity, $Z$, taken to be [Mg/H].
The observed abundance can be expressed
\begin{equation}\label{eq:xh}
    [\text{X}_j/\text{H}]_i = m_{ij} + \text{noise},
\end{equation}
where ``noise'' represents observational noise and/or other sources of intrinsic abundance scatter that are not included in this model. For detailed examples of a similar model with $K=2$, see Section 2 and Figures 2 and 3 of W22, where they demonstrate the vector addition and describe the process parameters for a few example stars.

In \name{}, we adopt the following set of assumptions: 

\paragraph{1. $K$ processes}
All elements on the periodic table are produced by a combination of nucleosynthetic processes such as CCSN, SNIa, AGB stars, and merging neutron stars \citep{johnsonja2020}. The majority of $\alpha$, light odd-$Z$, and Fe-peak elements (the elements observed by APOGEE) are dominantly produced by $K=2$ sources, with one being a prompt process or mix of prompt processes, and one being a delayed process or a mix of delayed processes. This is substantiated by theoretical yields \citep[e.g.,][]{anderson2019, rybizki2017} and past successful data-driven models (e.g., \citealp{ness2019}, G22, W22, \citealp{ting2022, ratcliffe2023}). In this paper we therefore assume that $K \geq 2$, though \name{} could in principle be implemented with $K=1$.

\paragraph{2. Linearity}
At every metallicity, the (linear) (X/H) abundances of a star can be expressed as a linear combination of $K$ processes.
These processes themselves will depend on metallicity, but a linear sum is sufficient to explain all element abundances at any overall metallicity.
Because different stars can get to their metallicities by different histories, and because detailed abundances beyond metallicity must matter at some level, the true enrichment mechanism is at least slightly nonlinear; thus this assumption
must be at least slightly wrong in detail.

\paragraph{3. Non-negativity}
All process vector components for all elements are non-negative and all process amplitudes are non-negative.
This assumption implies that the elements considered here are only produced, and not ever destroyed, by the $K$ processes (relative to hydrogen).
This makes the model similar to a non-negative matrix factorization (\citealt{nnmf, nnmf2}). In \name{}, this assumption is enforced by requiring that the process vector components and amplitudes are always greater than or equal to zero, such that
\begin{equation} \label{eq:constraint_zero}
    q_{k,j}^Z \geq 0 \; \forall \; Z,k,j \quad \text{and} \quad A_i^k \geq 0 \; \forall \; k,i ~.
\end{equation}

\paragraph{4. Mg production}
All Mg is produced in a prompt process and no other processes contribute to its production.
This is substantiated by theoretical yields where Mg is purely produced by prompt CCSN \citep[e.g.,][]{woosley1995, arnett1996, anderson2019, rybizki2017}.
This assumption (along with non-negativity) breaks a set of symmetries in the process space and makes the processes quasi-interpretable in terms of nucleosynthesis sources.
Because such a prompt process is likely dominated by CCSN \citep[e.g.,][]{andrews2017}, we label the first process with ``CC''. 

In \name{}, this assumption is enforced by fixing the Mg process vector components such that
\begin{equation}\label{eq:qcc_mg_solar}
    q_{{\rm CC, Mg}}^{\,Z} = 1, \quad q_{k>1,{\rm Mg}}^{\,Z} = 0
\end{equation}
at all metallicities. Equation~\ref{eq:qcc_mg_solar} also imposes that the Mg process is metallicity independent. 

\paragraph{5. Fe production}
Fe is produced through a combination of a prompt and delayed process. Because the delayed process is likely dominated by SNIa \citep[e.g.,][]{thielemann2002, andrews2017}, we label the delayed process with ``Ia''\footnote{While other enrichment channels with similar timescales may be included in the respective processes, the ``CC'' and ``Ia'' naming convention conforms to the choices in W22 and G22, and avoids the possible confusion of process numbers (1 and 2) with supernova type (II and Ia).}.
While the prompt process constraint (Mg) is grounded in nucleosynthesis theory, there is no equivalent nucleosynthesis fact to constrain the delayed process.
To break model degeneracies, we also fix the Fe process vector components such that 

\begin{equation}\label{eq:qcc_fe_solar}
    q_{{\rm CC, Fe}}^{\,Z} = 0.4, \quad q_{{\rm Ia, Fe}}^{\,Z} = 1 - q_{{\rm CC, Fe}}^{\,Z} \quad q_{k>2,{\rm Fe}}^{\,Z} = 0
\end{equation}
at all metallicities. 
This assumption places a star with purely prompt enrichment on the low metallicity $\mgfe$ plateau near 0.4 dex, in agreement with APOGEE observations but in contention with recent results from \citet{conroy2022} which place the $\mgfe$ plateau near 0.6 dex.
We explore the impact of different $\qccFe$ assumptions in Section~\ref{subsec:qccFe}.

\paragraph{6. Metallicity dependence}
We permit the process vector components for all elements other than Mg and Fe to float as a function of metallicity.
The variation is parameterized by a linear spline in log-process space, attached to a set of variable control points, knots, where the piecewise functions are joined. We assume that a particular set of 11 hard-coded knots between [Mg/H] of -0.8 and 0.6 are sufficient to capture the metallicity dependence. We choose knot number and location such that we capture the complex metallicity dependence of the abundance trends while maintaining a sufficient number of stars to fit with each linear component.

\paragraph{7. APOGEE abundances and uncertainties}
We assume that the APOGEE abundances and uncertainties can be used for this project.
This is not the same as assuming that they are correct, but rather that it is possible and useful to build an interpretable model to explain them. We describe the potential data systematics in Section~\ref{sec:data}.
For our purposes, we care mainly about the statistical observational errors rather than systematics that arise from imperfect modeling of the spectra such as NLTE effects, though differential systematics across the sample can artificially add abundance scatter. The actual derived values of $q_{k,j}^{Z}$ will be affected by systematic offsets in the abundances. We add a softening parameter $Q$ (Equation~\ref{eq:inflate_ivar}) to allow for the possibility that APOGEE observational errors are underestimated, or that there is intrinsic scatter around the \name{} predictions.

\paragraph{8. Robust likelihood function}
The observed value of [X/H] can be described as the $K$ process expected value plus observational noise and/or other sources of intrinsic abundance scatter, as described by Equation~\ref{eq:xh}.
The expression in this equation can be thought of as the key assumption underlying our likelihood function.
In detail the (negative two times the) log likelihood function is given by a chi-squared ($\chi^2$) objective
\begin{equation}\label{eq:chi2}
    \chi^2 = \sum_{ij}\frac{1}{\sigma_{ij}^2} \, (\xh_{ij} - m_{ij})^2 ~,
\end{equation}
where $1/\sigma_{ij}^2$ is the (robust; see below) inverse variance on measurement $ij$.

Because we don't want to be too drawn or influenced by outlier points, we don't use the observed errors $\sigma_{{\rm obs},ij}$ in the likelihood, but instead we soften them in the spirit of iteratively reweighted least squares (\citealt{irls}):
\begin{equation}\label{eq:inflate_ivar}
    \frac{1}{\sigma_{ij}^2} = \frac{Q^2/\sigma_{{\rm obs},ij}^2}{Q^2 + (\xh_{ij} - m_{ij})^2 /\sigma_{{\rm obs},ij}^2 },    
\end{equation}
where $Q$ is a softening parameter. Our results are largely insensitive to the choice of $Q$. We find that the predicted abundances of all elements change by less than 0.01 dex for Q between 1 and 10, so we choose to set $Q=5$. Very small Q values (e.g., $Q=.1$) will erase some of the abundance structure and produce poorer fits.

\paragraph{9. Implementation and optimization} 
With the above assumptions in place, the likelihood function can be optimized to a set of stellar abundances.
The model is initialized at the Mg and Fe process vector components from Equations~\ref{eq:qcc_mg_solar} and~\ref{eq:qcc_fe_solar}. It subsequently optimizes the process amplitudes (dubbed the $A$-step) using only Mg and Fe at fixed process vector components, and then optimizes the process vector components (dubbed the $q$-step) for all elements at fixed process amplitudes.
The $A$-step and $q$-step are alternated, repeating 48 rounds of optimization, in the $K=2$ case, and updating the best-fit parameters when the objective function improves. We find few differences in the best-fit parameters when we decrease the number of iterations to 32, indicating that the model quickly finds a good solution.  
In detail the optimizations are performed with a nonlinear $\chi^2$ minimization algorithm (Gauss--Newton nonlinear least-squares) from \texttt{jaxopt}\footnote{\url{https://jaxopt.github.io/}}.

\bigskip
\name{} mirrors the two-proccess model from prior work (G22, W22) but, unless otherwise noted, the assumptions are weaker, there is a likelihood function in play, and the implementation is more general.
In particular, we don't assume anything about the relationships between the process vector components and the morphologies of observed element-abundance ratio diagrams.

\section{Data}\label{sec:data}

In this paper, we employ stellar abundances from APOGEE DR17 \citep{abdurrouf2022}, part of the SDSS-IV \citep{majewski2017}. The APOGEE survey obtains high-resolution ($R\sim22,500$) near-infrared (IR) observations \citep{wilson2019} for stars in the Galactic disk, halo, bulge, and nearby satellites/streams. Observations are taken with two nearly identical spectrographs on the 2.5m Sloan Foundation telescope \citep{wilson2019} at Apache Point Observatory in New Mexico and the 2.5m du Pont Telescope \citep{bowen1973} at the Las Campanas Observatory in Chile. Spectral data are reduced and calibrated with the APOGEE data processing pipeline \citep{nidever2015}, after which stellar parameters and abundances are calculated with ASPCAP \citep[APOGEE Stellar Parameter and Chemical Abundance Pipeline;][]{holtzman2015, garcia2016}. See \citet[][DR16]{jonsson2020} and Holtzman et al. (in prep., DR17) for a more detailed description of APOGEE data reduction and analysis, and \citet{zasowski2013, zasowski2017}, \citet{beaton2021}, and \citet{santana2021} for a discussion of survey targeting.

APOGEE DR17 reports stellar parameters, including $\teff$ and $\logg$, as well as 20 elemental abundances: C, \ion{C}{1}, N, O, Na, Mg, Al, Si, S, K, Ca, \ion{Ti}{1}, \ion{Ti}{2}, V, Cr, Mn, Fe, Co, Ni, and Ce for 657,135 stars. In DR17, new spectral libraries \citep{hubeny2021} are generated using the Synspec code and incorporate NLTE corrections for Na, Mg, K, and Ca \citep{osorio2020}. Among the reported elements and ions, some are measured more precisely than others. We exclude Ti from our analysis as there are large differences between the abundances derived from the \ion{Ti}{1} and \ion{Ti}{2} lines \citep{jonsson2020}. We also exclude P and V, as the P abundances are measured from a few very weak spectra features and V abundances are one of the least precise and least accurate labels \citep{jonsson2020}. Both P and V display strong abundance artifacts and large scatter. Among the remaining elements, we note the following concerns: weak Na spectral features, large abundance scatter in S, significant systematic artifacts in Cr abundances at super-solar metallicities, potentially strong unaccounted-for NLTE effects on Mn abundances \citep{bergemann2019}, and large abundance scatter in Co and Ce. For a more detailed discussion of abundance systematics and their effects on population trends, see \citet{jonsson2020} and \citet{griffith2021a}.

For our stellar sample, we select a subset of APOGEE DR17 stars with the goal of minimizing statistical errors from poor observations and systematic errors from abundance trends with $\teff$ and/or $\logg$ while preserving a sufficient number of stars to conduct a meaningful statistical analysis across the Galactic disk. To remove poor quality data points, we require ASPCAP flags \texttt{STAR\_BAD} and \texttt{NO\_ASPCAP\_RESULT} equal zero. We only include stars from the main survey sample (\texttt{EXTRATARG} = 0), and use named abundances (\texttt{X\_FE}), as recommended by \citet{jonsson2020}. In addition to these quality cuts, we apply the following sample selection:
\begin{itemize}
\itemsep0em
    \item $\mgh > -0.75$
    \item S/N $\geq 200$
    \item $\logg = 1-3$ dex
    \item $\teff = 4000-5200$ K.
\end{itemize}
To eliminate red clump (RC) stars, which show abundance variations from the RGB sample \citep{vincenzo2021a}, we cross-match with and remove stars that appear in the APOGEE DR17 RC VAC\footnote{\url{https://www.sdss4.org/dr17/data_access/value-added-catalogs/?vac_id=apogee-red-clump-(rc)-catalog}} \citep{bovy2014}. 

\begin{figure*}[htb!]
    \centering
    \includegraphics[width=\textwidth]{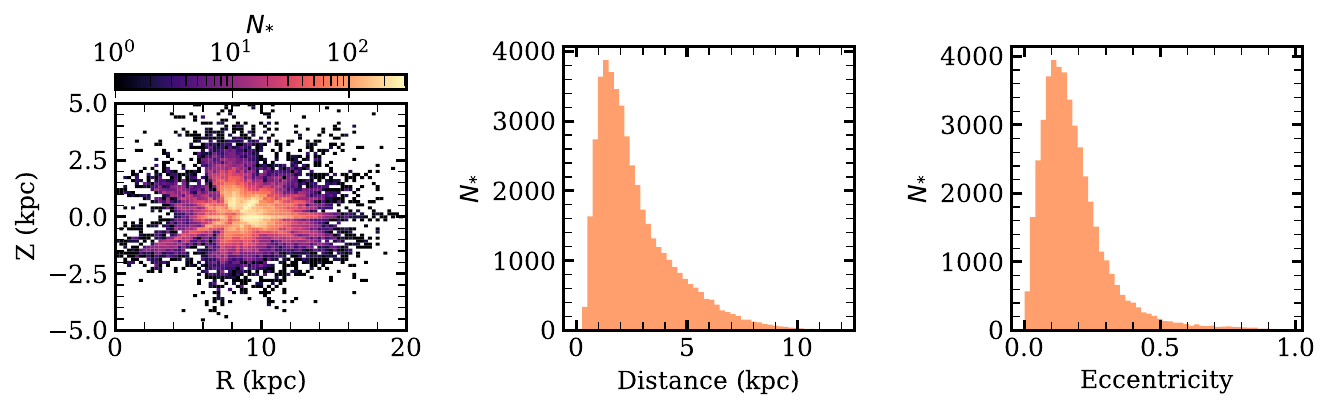}
    \caption{Left: distribution of our stellar sample in $Z$ (kpc) vs. $R$ (kpc) where (0,0) is the Galactic center. Center: distribution of stellar distances (kpc). Right: distribution of stellar eccentricity. Our stellar sample spans the Galactic disk, but the majority of our stars are within 3.5 kpc of the sun and have kinematics consistent with \textit{in situ} origin.}
    \label{fig:star_dist}
\end{figure*}

These cuts result in a sample of 48,659 stars that span the Galactic disk. We plot their $Z$ vs. $R$ location, as well as the distributions of distances and eccentricities in Figure~\ref{fig:star_dist}, taking distances and kinematics from \citep{queiroz2023}. While our stellar sample extends from Galactic center, to the outer disk, to the halo, the majority of our stars ($75\%$) are within 3.5 kpc of the sun. Further, $94\%$ of our stellar sample has an eccentricity less than 0.4, indicative of \textit{in situ} origin \citep[e.g.,][]{sales2009}. In this paper, we assume that the \name fits will be consistent across the Galactic disk, as the median high-Ia and low-Ia [X/Mg] vs. [Mg/H] abundance tends are insensitive to Galactic location (W19, \citealp{griffith2021a}).  While we fit a large sample of stars in this work, \name performs similarly when fitting smaller populations, down to 500 stars.

We present abundances for Mg, O, Si, S, Ca, C+N, Na, Al, K, Cr, Fe, Ni, Mn, Co, and Ce. In the analysis of each element, X, we drop stars with \texttt{X\_FE\_FLAG}. Ce abundances are flagged in the most stars, resulting in $\sim 700$ Ce labels being excluded. While the surface abundances of C and N differ from the stellar birth abundances for RGB stars due to the CNO processes and dredge-up events \citep{iben1965, shetrone2019}, the total C+N abundance remains constant. As in W22, we consider C+N as an element, taking [(C+N)/H] to be 
\begin{equation}
    [\text{C+N}/\text{H}] = \log_{10}(10^{\text{[C/H]}+8.39} + 10^{\text{[N/H]}+7.78}) - \log_{10}(10^{8.39} + 10^{7.78}),
\end{equation}
using logarithmic solar abundances for C (8.39) and N (7.78) from \citet{grevesse2007}. We further adopt the error on the [C/Fe] abundance as the error on [C+N/Fe], since C typically dominates in the abundance ratio.

\begin{figure*}[htb!]
    \centering
    \includegraphics[width=\textwidth]{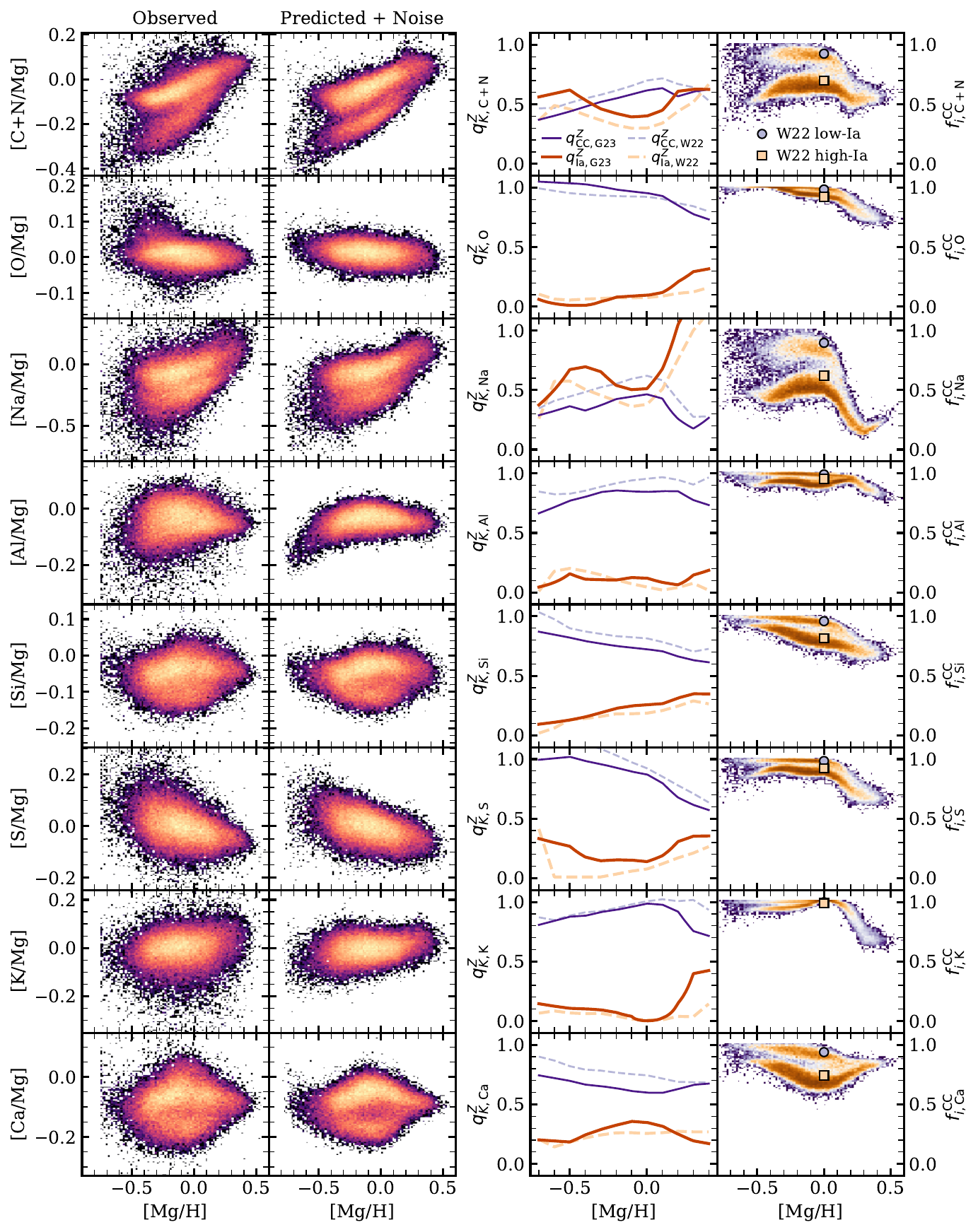}
    \caption{Abundance distributions and \name{} parameters for C+N, $\alpha$, and light odd-$Z$ elements. First column: observed abundance distributions in [X/Mg] vs. [Mg/H]. Second column: The predicted [X/Mg] vs. [Mg/H] abundance distribution of the fiducial model plus estimated noise. By comparing the first two columns we can evaluate the success of \name{} in reproducing the observed abundance distributions. Third column: process vector components $\qcc$ (thin, purple) and $\qIa$ (thick, orange) from this work (G23, solid, dark lines) and W22 (light, dashed lines). Overall offsets between the solid and dashed lines are driven largely by our normalization that places the [Mg/Fe] plateau at $+0.4$, rather than $+0.3$ in W22. Fourth column: distribution of fractional contribution from the prompt process ($\fcc$) predicted by the fiducial model. We plot the median $\fcc$ values of the low-Ia (orange square) and high-Ia (purple circle) populations in the solar metallicity bin from W22 for comparison. All density plots are logarithmically scaled.}
    \label{fig:all_param1}
\end{figure*}

\begin{figure*}[htb!]
    \centering
    \includegraphics[width=\textwidth]{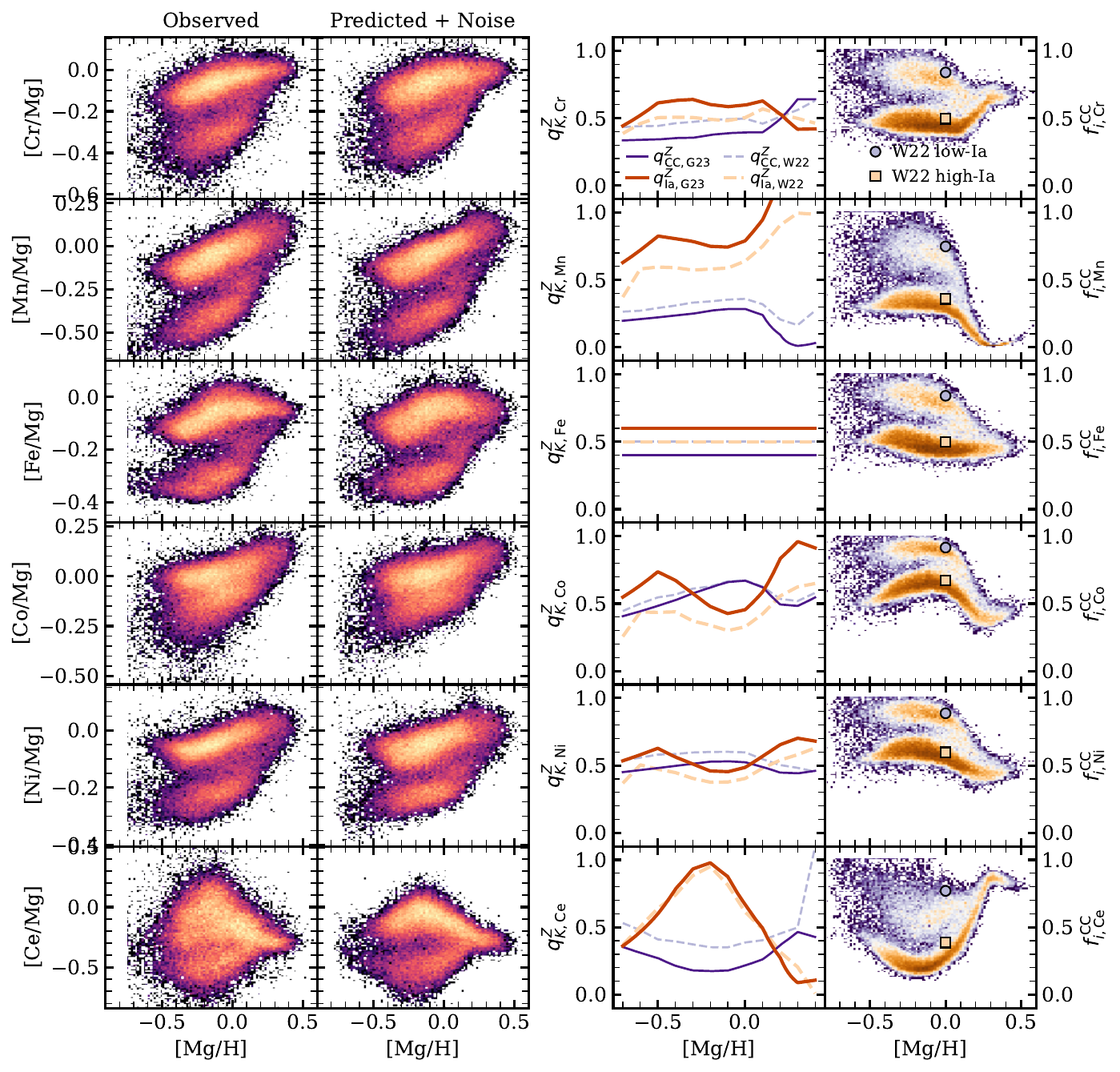}
    \caption{Same as Figure~\ref{fig:all_param1}, but for Fe-peak elements and Ce.}
    \label{fig:all_param2}
\end{figure*}

We plot the distributions of all abundances in [X/Mg] vs. [Mg/H] for our sample in the first column of Figures~\ref{fig:all_param1} and~\ref{fig:all_param2}. 

\section{The Fiducial Model} \label{sec:fiducial}

We fit the APOGEE sample with our fiducial model of $K=2$, such that
\begin{equation}\label{eq:m_ij}
    m_{ij} = \log_{10}(A_i^{\rm CC} \, q_{{\rm CC},j}^Z + A_i^{\rm Ia} \, q_{{\rm Ia},j}^Z)
\end{equation}
with the assumptions from Section~\ref{sec:model}. This fit produces process vector components $\qcc$ and $\qIa$ as a function of $\mgh$ for each element and process amplitudes $\Acc$ and $\AIa$ for each star. From the model parameters, we can calculate fractional contributions from each process as well as a full suite of predicted $K=2$ process abundances, shown in the second column of Figures~\ref{fig:all_param1} and~\ref{fig:all_param2}.

\subsection{Process Parameters and Fractional Contributions} \label{subsec:parameters}

We plot the process vector components as a function of $\mgh$ in the third column of Figures~\ref{fig:all_param1} and~\ref{fig:all_param2} and provide the values at the [Mg/H] knots in Tables~\ref{tab:qcc} and~\ref{tab:qIa}. The process vector components inform us about the relative contribution of prompt and delayed processes to the formation of the elements, as well as the metallicity dependence of the enrichment. By definition, $\qccFe=0.4$ at all metallicities. For Mg and Fe, we also require $\qcc + \qIa = 1$, implying $\qIaFe=0.6$. No such constraints are placed on other elements. We note that \name{} differs from the previous two-process models in this regard, as G22 and W22 require that the process vector components for all elements sum to 1 at solar metallicity.

\begin{deluxetable}{cccccccccccccccc}
    \tablecaption{Fiducial model $\qcc$ values at [Mg/H] knot values for each element.
    \label{tab:qcc}}
    \tablehead{
\colhead{[Mg/H]} &	\colhead{C+N} &	\colhead{O} &	\colhead{Na} &	\colhead{Mg} &	\colhead{Al} &	\colhead{Si} &	\colhead{S} &	\colhead{K} &	\colhead{Ca} &	\colhead{Cr} &	\colhead{Mn} &	\colhead{Fe} &	\colhead{Co} &	\colhead{Ni} &	\colhead{Ce}} 
    \startdata
-0.8 &	0.34 &	1.06 &	0.26 &	1.0 &	0.62 &	0.90 &	0.98 &	0.78 &	0.77 &	0.33 &	0.19 &	0.4 &	0.37 &	0.44 &	0.40 \\
-0.5 &	0.44 &	1.04 &	0.36 &	1.0 &	0.77 &	0.82 &	1.02 &	0.88 &	0.70 &	0.34 &	0.22 &	0.4 &	0.48 &	0.48 &	0.27 \\ 
-0.4 &	0.48 &	1.03 &	0.33 &	1.0 &	0.80 &	0.79 &	0.98 &	0.88 &	0.67 &	0.35 &	0.24 &	0.4 &	0.52 &	0.50 &	0.21 \\
-0.3 &	0.52 &	1.01 &	0.37 &	1.0 &	0.84 &	0.77 &	0.96 &	0.92 &	0.65 &	0.35 &	0.25 &	0.4 &	0.58 &	0.51 &	0.18 \\
-0.2 &	0.55 &	0.98 &	0.42 &	1.0 &	0.86 &	0.75 &	0.93 &	0.93 &	0.64 &	0.38 &	0.27 &	0.4 &	0.62 &	0.53 &	0.17 \\
-0.1 &	0.58 &	0.97 &	0.45 &	1.0 &	0.85 &	0.74 &	0.90 &	0.96 &	0.61 &	0.39 &	0.29 &	0.4 &	0.66 &	0.53 &	0.18 \\
0.0 &	0.62 &	0.95 &	0.46 &	1.0 &	0.84 &	0.72 &	0.87 &	0.99 &	0.60 &	0.39 &	0.29 &	0.4 &	0.67 &	0.52 &	0.22 \\
0.1 &	0.64 &	0.93 &	0.43 &	1.0 &	0.85 &	0.71 &	0.80 &	0.98 &	0.60 &	0.40 &	0.24 &	0.4 &	0.62 &	0.49 &	0.26 \\
0.2 &	0.57 &	0.85 &	0.26 &	1.0 &	0.85 &	0.66 &	0.68 &	0.92 &	0.63 &	0.49 &	0.09 &	0.4 &	0.49 &	0.45 &	0.37 \\
0.3 &	0.61 &	0.78 &	0.18 &	1.0 &	0.78 &	0.63 &	0.62 &	0.76 &	0.66 &	0.64 &	0.01 &	0.4 &	0.48 &	0.44 &	0.47 \\
0.6 &	0.66 &	0.65 &	0.61 &	1.0 &	0.66 &	0.58 &	0.49 &	0.64 &	0.70 &	0.64 &	0.28 &	0.4 &	0.70 &	0.50 &  0.36 \\ 	
\enddata
\end{deluxetable}

\begin{deluxetable}{cccccccccccccccc}
    \tablecaption{Fiducial model $\qIa$ values at [Mg/H] knot values for each element.
    \label{tab:qIa}}
    \tablehead{
\colhead{[Mg/H]} &	\colhead{C+N} &	\colhead{O} &	\colhead{Na} &	\colhead{Mg} &	\colhead{Al} &	\colhead{Si} &	\colhead{S} &	\colhead{K} &	\colhead{Ca} &	\colhead{Cr} &	\colhead{Mn} &	\colhead{Fe} &	\colhead{Co} &	\colhead{Ni} &	\colhead{Ce}} 
    \startdata
-0.8 &	0.54 &	0.15 &	0.28 &	0.0 &	0.02 &	0.08 &	0.37 &	0.17 &	0.21 &	0.38 &	0.55 &	0.6 &	0.47 &	0.49 &	0.28 \\
-0.5 &	0.62 &	0.01 &	0.67 &	0.0 &	0.16 &	0.13 &	0.27 &	0.11 &	0.18 &	0.61 &	0.83 &	0.6 &	0.74 &	0.63 &	0.60 \\
-0.4 &	0.54 &	0.01 &	0.70 &	0.0 &	0.11 &	0.15 &	0.18 &	0.10 &	0.24 &	0.63 &	0.81 &	0.6 &	0.67 &	0.56 &	0.78 \\
-0.3 &	0.46 &	0.04 &	0.66 &	0.0 &	0.11 &	0.19 &	0.15 &	0.09 &	0.29 &	0.64 &	0.79 &	0.6 &	0.58 &	0.51 &	0.93 \\
-0.2 &	0.42 &	0.08 &	0.54 &	0.0 &	0.10 &	0.23 &	0.15 &	0.07 &	0.33 &	0.60 &	0.75 &	0.6 &	0.48 &	0.46 &	0.98 \\
-0.1 &	0.39 &	0.09 &	0.50 &	0.0 &	0.13 &	0.25 &	0.15 &	0.04 &	0.36 &	0.58 &	0.74 &	0.6 &	0.43 &	0.45 &	0.88 \\
0.0 &	0.40 &	0.10 &	0.51 &	0.0 &	0.12 &	0.26 &	0.14 &	0.00 &	0.35 &	0.60 &	0.79 &	0.6 &	0.45 &	0.49 &	0.67 \\
0.1 &	0.45 &	0.12 &	0.68 &	0.0 &	0.08 &	0.27 &	0.19 &	0.03 &	0.32 &	0.63 &	0.95 &	0.6 &	0.59 &	0.57 &	0.49 \\
0.2 &	0.61 &	0.21 &	1.05 &	0.0 &	0.06 &	0.31 &	0.31 &	0.15 &	0.25 &	0.53 &	1.26 &	0.6 &	0.83 &	0.65 &	0.28 \\
0.3 &	0.63 &	0.29 &	1.32 &	0.0 &	0.15 &	0.35 &	0.35 &	0.40 &	0.19 &	0.42 &	1.46 &	0.6 &	0.96 &	0.70 &	0.09 \\
0.6 &	0.62 &	0.37 &	1.13 &	0.0 &	0.30 &	0.34 &	0.36 &	0.48 &	0.13 &	0.43 &	1.40 &	0.6 &	0.83 &	0.64 &	0.16\\
\enddata
\end{deluxetable}

In the fourth column of Figures~\ref{fig:all_param1} and~\ref{fig:all_param2}, we plot the distribution of fractional contributions from the prompt process ($\fcc$) to each element, where
\begin{equation}\label{eq:fcc}
    \fcc = \frac{\Acc \, \qcc}{\Acc \, \qcc + \AIa \, \qIa}.
\end{equation}
We generally find that the distributions are bimodal, like the observed abundance patterns, as the high-Ia and low-Ia populations have differing fractional contributions from prompt and delayed sources.

We find that the $\alpha$-elements (O, Si, S, Ca) are best fit with $\qcc$ and $\fcc > 0.5$ at all metallicities. This is in agreement with theoretical prediction that $\alpha$-elements are dominated by prompt CCSN production \citep[e.g.][]{andrews2017}. O, a Mg-like element theoretically purely produced in prompt CCSN, shows $\fccX{O}$ near 1 from $\mgh=-0.75$ to solar. At supersolar metallicity, the delayed process contributes to O production, driving the $\fccX{O}$ value down to $\sim 0.8$ at $\mgh=0.4$. S behaves like O, with almost entirely prompt production up to solar metallicity, after which delayed enrichment contributes more significantly. Conversely, we find that Si and Ca are best fit with prompt and delayed enrichment at all metallicities, though the prompt process always dominates. For Si, the delayed process appears to increase linearly with $\mgh$, while the Ca delayed enrichment increases from $\mgh$ of $-0.75$ to $-0.1$ and then decreases from $\mgh$ of $-0.1$ to $0.5$.

The process vector components of light odd-$Z$ elements Al and K resemble those of the $\alpha$-elements, such as S. Both exhibit $\qcc$ and $\fcc$ near 1 through solar metallicity, with an increase in $\qIa$ and downturn in $\fcc$ at supersolar metallicities (especially for K). The behavior of the Na process vector components is more complex, with peaks and troughs in $\qIaX{Na}$. We find that Na has the strongest contributions from the delayed process of all $\alpha$ and light odd-$Z$ elements, with $\qIaX{Na} \gtrsim 0.5$ at almost all values of $\mgh$ and $\fccX{Na} < 0.3$ at $\mgh > 0$. The strong delayed contribution to Na is in agreement with findings of W22 and G22, and in tension with theoretical yields \citep[e.g.][]{andrews2017, rybizki2017}.

Unlike $\alpha$ and light odd-$Z$ elements whose delayed production is dominated by SNIa, C and N are thought to be promptly produced in CCSN with additional delayed enrichment from AGB stars \citep[e.g.][]{andrews2017}. We find that the prompt and delayed processes both contribute significantly, and nearly equally, across our stellar sample. Though theoretical N yields from AGB stars have a strong metallicity dependence \citep{karakas2010, ventura2013, cristallo2015, johnson2022}, we observe only a slight positive metallicity dependence in $\qccX{C+N}$ and a shallow dip in $\qIaX{C+N}$. We find a population of stars with $\fccX{C+N}$ near 0.9 and a population near 0.4. 

The Fe-peak elements (Cr, Mn, Fe, Co, Ni) are thought to be produced through prompt CCSN production and delayed SNIa production \citep[e.g.][]{andrews2017}. By construction, $\qccFe=0.4$ and $\qIaFe=0.6$ at all metallicities. This produces a bimodal distribution in $\fccX{Fe}$ similar to that observed in abundance space. Because of our choice of $\qccFe$, only a few stars have $\fccX{Fe}=1$ (see Section~\ref{subsec:qccFe}). We instead observe a population with $\fccX{Fe}$ near 0.8 and a population near 0.4. The process vector components and $\fccX{Fe}$ distribution for Cr and Ni strongly resemble those of Fe. All three elements have even atomic numbers. At supersolar metallicity, we find that the prompt process dominates Cr production, resulting in an upturn in $\fcc$. Conversely, Ni displays a dominant, and increasing, delayed process vector component at supersolar metallicities. The process vector components for Mn and Co (odd atomic numbers) show a complex metallicity dependence, more resembling that of Na. Both elements display a strong delayed process, with the $\qIaX{Mn} > 0.5$ at all metallicities and $> 1$ for $\mgh > 0.1$. Mn is the only element for which $\fccX{Mn}$ decreases to 0 for $\mgh \gtrsim 0.2$. 

Finally, we find that the delayed process dominates Ce production at intermediate metallicity, with $\qIaX{Ce}$ increasing up to $\mgh \approx -0.2$ and then decreasing to nearly 0 at $\mgh \approx 0.3$. The $\fccX{Ce}$ values are clustered near 0.25 around $\mgh$ of 0.2, then increase such that the abundances are almost entirely dominated by prompt enrichment at high metallicity.

\begin{figure*}[htb!]
    \centering
    \includegraphics[width=\textwidth]{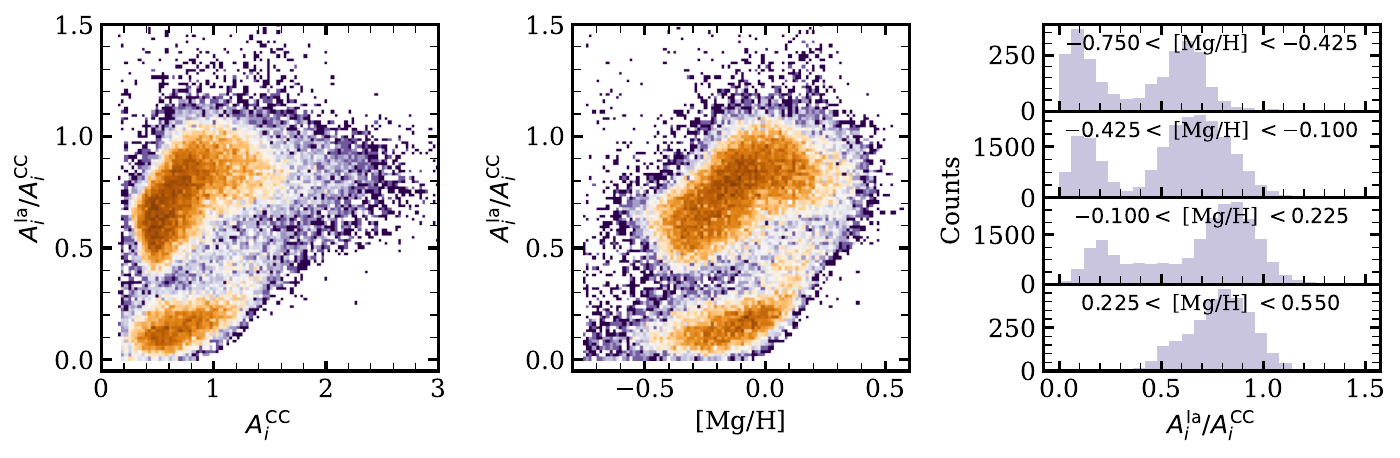}
    \caption{Left: distribution of $\AIa/\Acc$ vs. $\Acc$ for the fiducial model. This plot is similar to a [Fe/Mg] vs. [Mg/H] distribution where $\AIa/\Acc$ is a proxy for [Fe/Mg] and $\Acc$ is a proxy for [Mg/H]. Note that $\Acc = 1$ corresponds to $\mgh = 0$. Center: distribution of $\AIa/\Acc$ vs. $\mgh$. In the left and center panels, we can clearly see the bimodality to high values of $\Acc$ and $\mgh$. Both density plots are logarithmically scaled. Right: distribution of $\AIa/\Acc$ for ranges of [Mg/H] with metallicity bin increasing from top to bottom. [Mg/H] bins are of width 0.325 dex and span $-0.75$ to 0.55.}
    \label{fig:As}
\end{figure*}

In addition to process vector components, each star is fit with a prompt and delayed process amplitude, $\Acc$ and $\AIa$ respectively (Table~\ref{tab:As}). All elemental abundances are used in the calculation of these amplitudes, so they can be interpreted as ``de-noised'' abundance labels that suppress observational scatter by averaging over elements via the data-driven model. The value of $\Acc$ traces the metallicity (specifically $\mgh$) while the ratio of $\AIa/\Acc$ traces the $\femg$ abundance. In the left panel of Figure~\ref{fig:As} we plot the distribution of $\AIa/\Acc$ vs. $\Acc$. We find a bimodal distribution, similar to the Tinsley-Wallerstein diagram ($\mgfe$ vs. $\feh$, \citealp{wallerstein1962, tinsley1979, tinsley1980}), as was found in W22 and G22. We stress that the presence of the abundance bimodality was not fed into our model, and yet it is recovered in the best-fit process amplitudes. The stars with larger $\AIa/\Acc$ values correspond to the high-Ia population, and those with low $\AIa/\Acc$ correspond to the low-Ia population. While in the Tinsley-Wallerstein diagram the two populations blend together at high metallicity, they are more distinguishable in our amplitude space. We plot $\AIa/\Acc$ vs. $\mgh$ in the center panel of Figure~\ref{fig:As}. The high-Ia and low-Ia populations are clearly separable through $\mgh$ of 0.4. This is further shown through the $\AIa/\Acc$ distributions in the right panel of Figure~\ref{fig:As} for [Mg/H] bins of $-0.75$ to $-0.425$, $-0.425$ to $-0.1$, $-0.1$ to 0.225, and 0.225 to 0.55. The three lowest metallicity bins display a bimodal distribution and the highest metallicity bin is dominated by high-Ia stars.

\begin{deluxetable}{cccc}
    \tablecaption{Fiducial model $\Acc$ and $\AIa$ values for our stellar sample.
    \label{tab:As}}
    \tablehead{
\colhead{APOGEE ID} &	\colhead{[Mg/H]} & \colhead{$\Acc$} &	\colhead{$\AIa$}} 
\startdata
2M00000546+6152107 &	-0.20 &	0.61 &	0.48\\
2M00000866+7122144 &	-0.10 &	0.82 &	0.69\\
2M00001328+5725563 &	0.04 &	1.09 &	1.02\\
2M00001653+5540107 &	0.06 &	1.13 &	0.36\\
2M00001717+6147500 &	-0.23 &	0.60 &	0.39\\
... & ... & ... & ...\\
\enddata
\tablecomments{Full table available online.}
\end{deluxetable}

With the optimized process parameters in hand, we can use Equation~\ref{eq:xh} to calculate predicted abundances for the fiducial model---the abundances our stellar population would have if the model assumptions are correct and only one prompt and one delayed process contribute. To simulate observational noise, we add an error drawn from a Gaussian distribution with $\sigma$ equal to the reported error on each abundance for each star. In Figure~\ref{fig:all_param1} and~\ref{fig:all_param2} we plot the predicted abundances plus estimated noise in the second columns. These distributions can be compared to the observed abundance distribution in the first columns.

Overall, the fiducial model successfully reproduces the observed abundance distributions. It is capable of capturing metalicity dependences and bimodality. The predicted abundances plus estimated noise are not, however, able to reproduce the observed abundance scatter. This is especially noticeable for O, C+N, Na, Al, K, Co, and Ce. For these elements the scatter in the observed abundance distribution is larger than in the predicted distribution, suggesting that the APOGEE observational scatter is underestimated, that there are $\teff$ or $\logg$ dependent abundance trends (e.g., \citealp{griffith2021a}, W22), or that the $K=2$ model is insufficient--a likely case for elements produced by AGB stars, such as C+N and Ce. The model performs similarly well when the population is downsampled to 5000, 1000, and 500 stars, though the number of nodes has to be decreased from 11 to 7.

\subsection{Comparing to W22}\label{subsec:w22}

As discussed in Sections~\ref{sec:intro} and~\ref{sec:model}, \name{} is based upon the two-process model developed in W19 and W22, but with increased flexibility, minimal normalization, and no forced dependence upon the [Fe/Mg] vs. [Fe/H] bimodality or population abundance trends. Further, the \name{} utilizes all stellar abundances in the optimization of $\Acc$ and $\AIa$, whereas only Mg and Fe are used in W19 and only Mg, O, Si, Ca, Fe, and Ni in W22. 

In the fiducial model, we adopt $K=2$, as in W22, but assume $\qccFe = 0.4$, 0.1 dex lower than the $\qccFe$ value assumed in W22. In practice, this moves the implied ``pure'' CCSN enrichment plateau from $\femg=-0.3$ to $\femg=-0.4$ (though the W22 plateau value is determined \textit{after} they apply a global offset of +0.05 to all $\femg$ abundances). Because our model is non-negative, it requires a lower $\qccFe$ to correctly model the stars on the $\femg$ plateau, whereas W22 assigns stars with $\femg < -0.3$ negative $\AIa$ values.

While our stellar samples and model assumptions differ, we plot the W22 $\qcc$ and $\qIa$ vector components as well as the W22 solar metallicity $\fcc$ values of Figures~\ref{fig:all_param1} and~\ref{fig:all_param2} for comparison with our fiducial model. We generally observe similar behavior between \name{} and W22. Our $\qIa$ vector components tend to be $\sim 0.1$ greater than those of W22 for elements with significant delayed contributions because of our differing $\qccFe$ assumptions. The metallicity dependencies agree for most elements, with small variations at the high-$\mgh$ end for O, Al, K, and Ce. We also see good agreement between the \name{} and W22 solar metallicity $\fcc$ values, with the W22 points slightly offset to larger values for elements with significant delayed contributions.

\begin{figure*}[htb!]
    \centering
    \includegraphics[width=\textwidth]{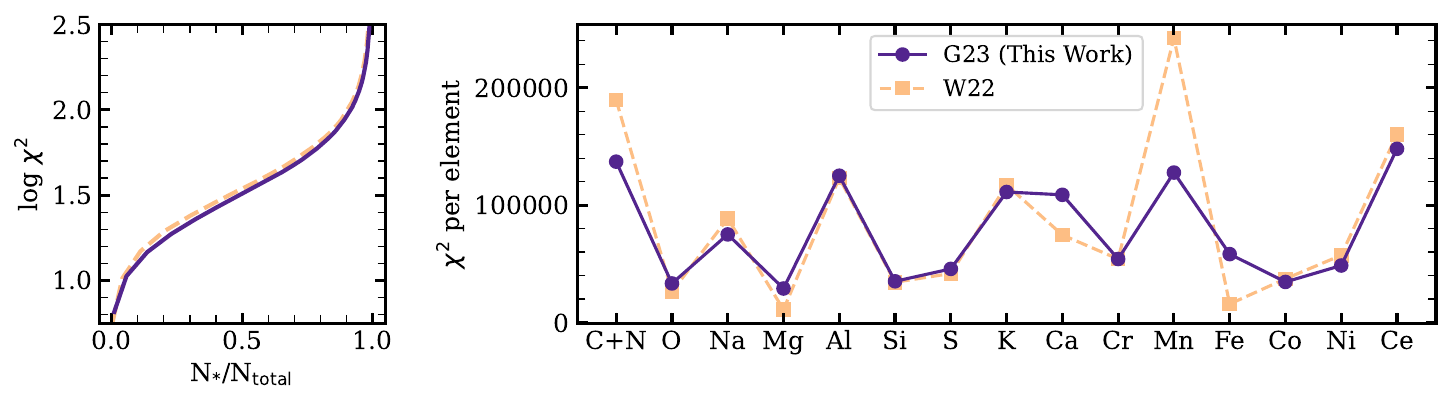}
    \caption{Left: cumulative distribution of $\log_{10}(\chi^2)$ for W22 (dashed orange line) and our fiducial model (G23, solid purple line). Right: $\chi^2$ per element for the same model fits with elements ordered by atomic number. Overall, our model has a smaller cumulative $\chi^2$ than the previous two-process model and better predicts the abundances of C+N, Na, K, Mn, Co, Ni, and Ce.}
    \label{fig:comp_w22}
\end{figure*}

To compare the accuracy of the models' abilities to reproduce the observed abundances, we identify a subset of $\sim 23,000$ stars in both our sample and the W22 sample. We calculate the predicted abundances for each star under \name{} and the two-process model, then determine the $\chi^2$ value of the fits for each star (summing over the elements) and for each element (summing over the stars). We plot the cumulative stellar $\log(\chi^2)$ distribution and the total $\chi^2$ for each element in the right and left panels, respectively, of Figure~\ref{fig:comp_w22}. It is important to note that in the calculation of the W22 model residuals, we do not apply the temperature corrections discussed in Section 5.1 of W22. 

We find that, overall, the $\chi^2$ decreases between the W22 two-process model and our $K$-process model, an indication that we better predict all of a star's abundances. When looking at each element individually, we find that we better predict C+N, Na, K, Ni, Mn, Co, and Ce, with major improvements to C+N and Mn. Our fiducial model is significantly worse at predicting Mg, Ca, and Fe than the W22 model, three of the six elements that W22 employ to fit the process amplitudes. Because \name{} uses all elements in its optimization, Mg, Ca, and Fe are effectively de-weighted relative to the W22 model, while C+N and Mn influence the model parameters. If we re-fit \name{} using only the Mg, O, Si, Ca, Fe, and Ni abundances in the $A$-step (as in W22), we find that \name{} and the two-process model predict the abundances of all elements but Mn with similar accuracy, and that the two-process model better predicts Mn. Our fiducial model's success in predicting C+N and Mn is likely attributable to the inclusion of the elements in the $A$-step. The choice to include all elements or a subset of elements in the fits should be considered when implementing \name{}. If searching for stars with anomalous abundances of element X relative to the expected abundances of others, one may want to exclude X from the $A$-step.

We note that our fiducial model is fit to a stellar sample that spans a wider range of $\teff$ and $\logg$ than the W22 sample. If we repeat our analysis on the W22 stellar sample with $\qccFe=0.5$ we almost perfectly recover the W22 process vector components, with small deviations at $\mgh>0.1$, and more substantially improve upon the stellar and elemental $\chi^2$ values. Most notably, \name{} is better able to predict the abundances of stars with $\mgh>0$, where the high-Ia and low-Ia sequences blend together and the W22 categorization of high-Ia and low-Ia stars may be incorrect.

\section{Variations Away from the Fiducial Model} \label{sec:variations}

\subsection{Impact of Fe Process Assumptions} \label{subsec:qccFe}

While \name{} is flexible, we still include some quantitative assumptions, which we make to break exact model degeneracies (see Section~\ref{sec:model}).
Specifically, for each process, we choose one element to assign a ``known'' process vector component at all metallicities.
These are Mg and Fe in the $K=2$ case.
Specifically, for the prompt process, we ground our assumption in the nucleosynthetic theory that Mg is a pure CCSN element \citep[e.g.,][]{andrews2017}.
Unfortunately, there is no comparable pure SNIa element, nor is there an element for which we know the relative CCSN/SNIa ratio.
In order to break an exact degeneracy between the prompt and delayed processes, we choose to fix the Fe process vector components, which effectively makes assumptions about the exact fractional contribution of CCSN and SNIa (or prompt and delayed processes) to Fe enrichment.

\begin{figure*}[htb!]
    \centering
    \includegraphics[width=\textwidth]{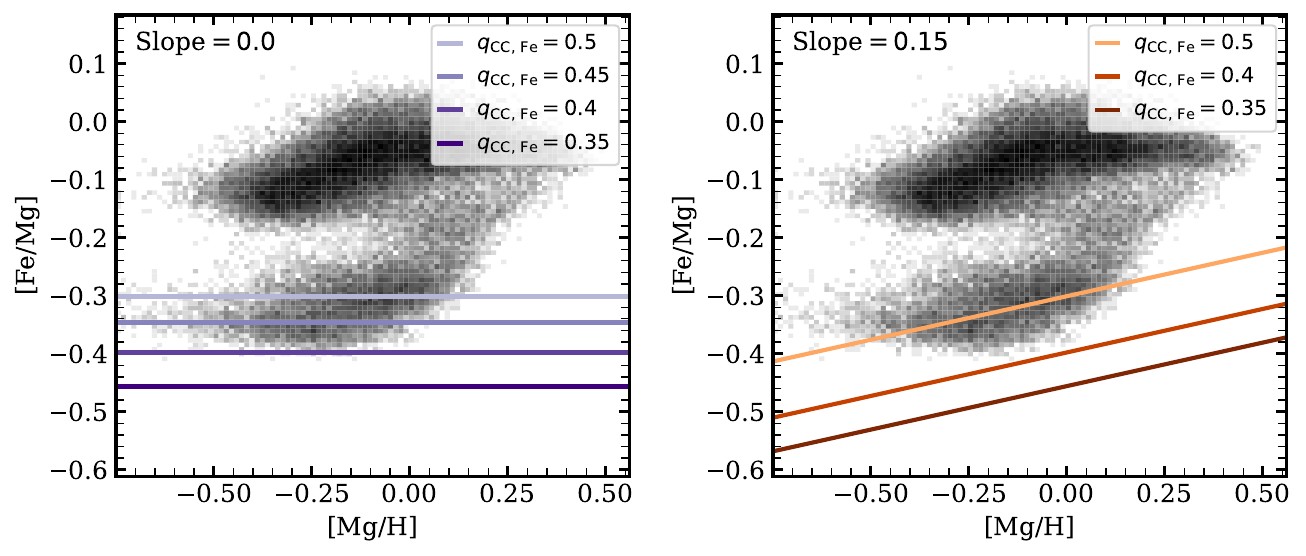}
    \caption{Minimum [Fe/Mg] value attainable for $\qccFe$ and $\dqccFe$ assumptions. Left: for $\dqccFe=0$ with $\qccFe=0.5$, 0.45, 0.4, and 0.35 (light to dark purple). Right: for $\dqccFe=0.15$ with $\qccFe=0.5$, 0.4, 0.35 (light to dark orange).}
    \label{fig:qccFe_FeMg}
\end{figure*}

In the fiducial model, we choose $\qccFe = 0.4$ as this parameter choice is able to reproduce the observed [Fe/Mg] vs. [Mg/H] abundance distribution, as discussed in Section~\ref{subsec:parameters}. This choice impacts the predicted abundances as well as the implied $\fcc$ values of each star. Because our model is non-negative, the choice of $\qccFe$ sets the minimum [Fe/Mg] value attainable by our model ($\log_{10}(\qccFe)$). In this Section, we explore the implications of different $\qccFe$ assumptions, varying the zero point and metallicity dependence ($\dqccFe$). In Figure~\ref{fig:qccFe_FeMg}, we plot the minimum [Fe/Mg] as a function of $\mgh$ for the $\qccFe$ and $\dqccFe$ parameters we explore. With a $\dqccFe=0.0$, we choose $\qccFe=0.5$ (the W22 value), 0.45 (approximate plateau value at $\mgh=-0.75$), 0.4 (the fiducial value that skirts the edge of the distribution), and 0.35 (captures almost all stars). With a $\dqccFe=0.15$, which roughly matches the slope of the low-Ia sequence at intermediate metallicity, we choose $\qccFe=0.5$ (passes through the center of the low-Ia density), 0.4 (skirts the edge of the distribution), and 0.35 (captures almost all stars).

\begin{figure*}[htb!]
    \centering
    \includegraphics[width=\textwidth]{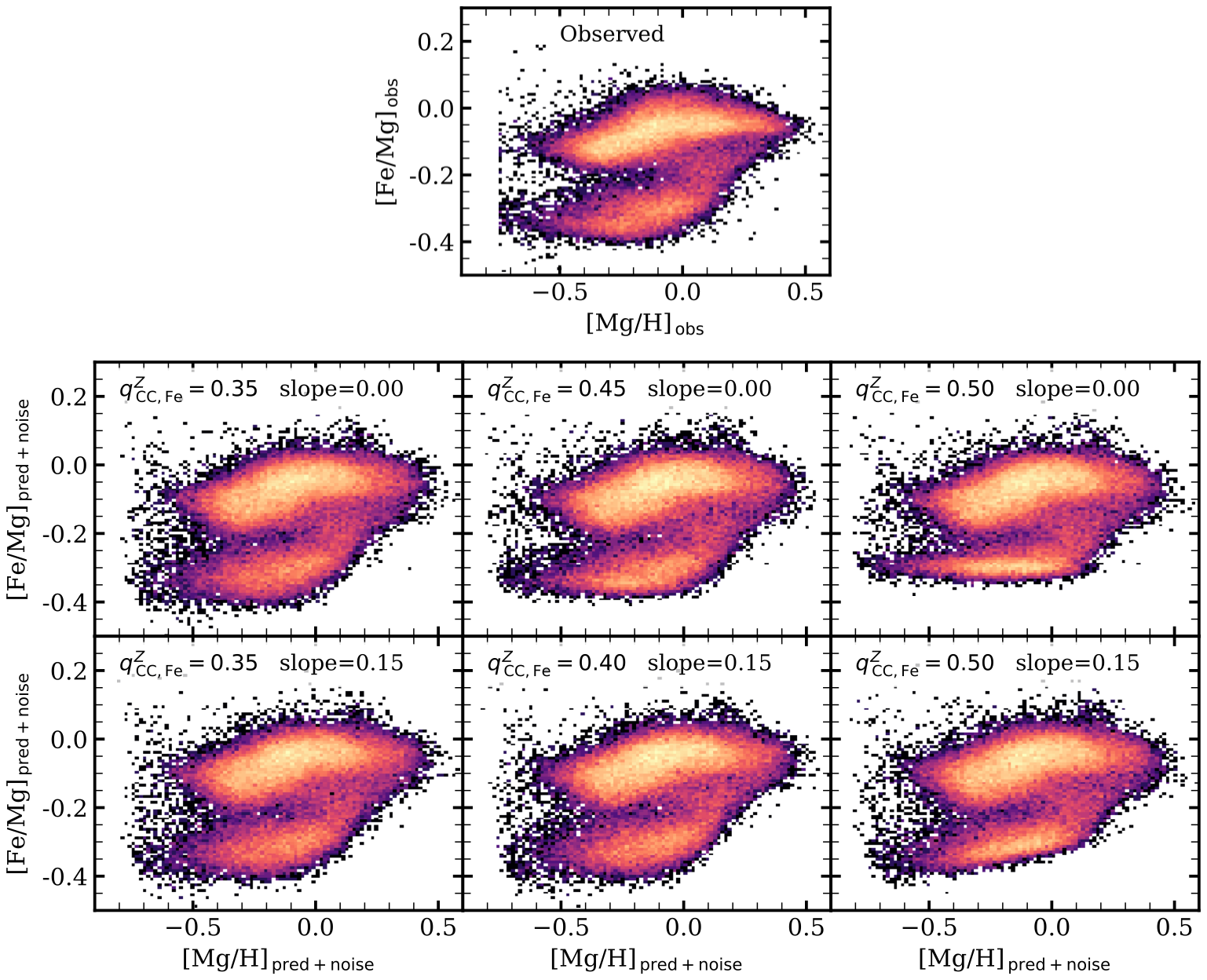}
    \caption{Top: observed [Fe/Mg] vs. [Mg/H] abundance distribution. Bottom: predicted [Fe/Mg] vs. [Mg/H] abundance distributions plus estimated noise for models with $\dqccFe=0.0$ (middle row) and $\dqccFe=0.15$ (bottom row). The $\qccFe$ value increase from left to right from 0.35 to 0.5. All density plots are logarithmically scaled. Note that the center panels have different $\qccFe$ values (middle: 0.45, bottom: 0.4) and that the predicted abundance distribution for $\qccFe=0.4$ and $\dqccFe=0.0$ can be found in Figure~\ref{fig:all_param2}. The $\qccFe$ and $\dqccFe$ parameters are listed in each panel. By comparing the lower panels to the observed distribution (top) we can evaluate the success or failure of each model; the low-Ia population is most sensitive to the Fe assumptions.}
    \label{fig:qccFe_FeMgpred}
\end{figure*}

We repeat the optimization of the \name{} with $K=2$ and these $\qccFe$ assumptions. In Figure~\ref{fig:qccFe_FeMgpred}, we plot the resulting predicted abundance distributions plus estimated noise alongside the observed distribution. We do not plot the prediction for the fiducial model ($\qccFe=0.4$, $\dqccFe=0.0$), as this is shown in Figure~\ref{fig:all_param2}. Both models with $\qccFe=0.5$ fail to reproduce the shape and width of the low-Ia abundance distribution. They instead predict a much thinner sequence that is flat for $\dqccFe=0.0$ or slightly inclined for $\dqccFe=0.15$. The model with $\qccFe=0.45$ and $\dqccFe=0.0$ is better, but still predicts a low-Ia abundance distribution that is too thin, flat, and dense. The other four models ($\qccFe=0.35$ and $\qccFe=0.4$ with $\dqccFe=0.0$ and $\dqccFe=0.15$) predict an abundance distribution that strongly resembles the observed. There are minor differences in the low [Fe/Mg] and low [Mg/H] region, but it is difficult to tell by eye which model is best.

To better assess the goodness of fit of each model, we calculate the average $\chi^2$ value per star. The models with $\qccFe$ of 0.5 and 0.45, regardless of $\dqccFe$, have an average $\chi^2$ per star $>90$ while the models with $\qccFe$ of 0.4 and 0.35 have an average $\chi^2$ per star $< 55$. In both the metallicity independent and metalicity dependent cases, the models with $\qccFe=0.4$ have the lowest average $\chi^2$ per star, at 54.54 and 54.47, respectively, though the models with $\qccFe=0.35$ have a $\chi^2$ that is only greater by $\sim0.1$. Of the seven models explored here, the case with $\qccFe=0.4$ and $\dqccFe=0.15$ has the lowest average $\chi^2$ per star, indicating that the Fe abundances are best fit by a metallicity dependent prompt process. Introducing this metallicity dependence subtly changes the shape of the predicted low-Ia distribution in a way that achieves better agreement with APOGEE observations.

\begin{figure*}[htb!]
    \centering
    \includegraphics[width=\textwidth]{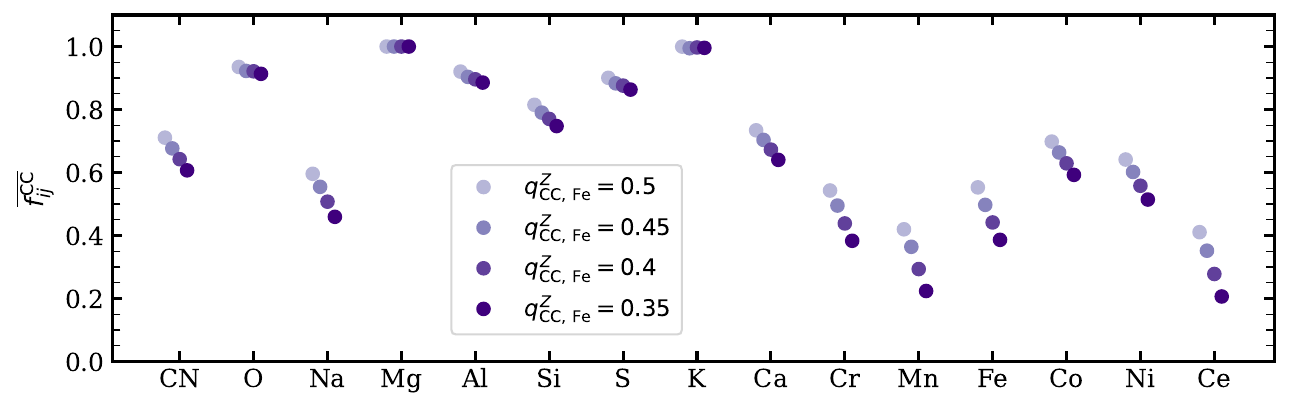}
    \caption{Elemental median values of $\fcc$ at solar metallicity for the low-Ia population for \name{} with $\qccFe=0.35$ (darkest purple) to 0.5 (lightest purple) and $\dqccFe=0.0$. Elements are ordered by atomic number. The median $\fcc$ changes most dramatically for elements with strong delayed contributions.}
    \label{fig:qccFe_fcc}
\end{figure*}

Though the $\qccFe= 0.4$ and 0.35 models are similar in terms of their goodness of fit, their nucleosynthesis implications are different. In Figure~\ref{fig:qccFe_fcc}, we plot the median value of $\fcc$ (Equation~\ref{eq:fcc}) for the low-Ia population at solar metallicity ($-0.05 < \mgh < 0.05$), where low-Ia stars are defined by
\begin{equation}\label{eq:lowIa}
\begin{cases}
\mgfe > 0.12 - 0.13 \, \feh,    & \feh<0 \cr
\mgfe > 0.12,               & \feh>0, \cr
\end{cases}
\end{equation}
as in W19, W22, and G22.
We only show the median $\fcc$ values for the models with $\dqccFe=0.0$ as the solar metallicity median $\fcc$ values for the metallicity dependent models are almost identical for matching values of $\qccFe$. We find that the choice of $\qccFe$ has little impact on the median $\fcc$ values of elements dominated by CCSN enrichment (e.g., O, Al, S, K). As the delayed contribution increases, the median elemental $\fcc$ values decrease more significantly with decreasing $\qccFe$. The choice of $\qccFe$ most impacts the median $\fcc$ values for Na, Cr, Fe, Mn, and Ce, with the median $\fcc$ for Mn decreasing from 0.42 for $\qccFe=0.5$ to 0.22 for $\qccFe=0.35$. Because the $\qccFe$ value sets the prompt enrichment plateau, a lower $\qccFe$ model implies a lower $\fcc$ value. 

While the high $\qccFe$ model can likely be ruled out due to poorness of fit, the true $\qccFe$ value and its metallicity dependence are unknown. It is therefore important not to over-interpret the specific $\fcc$ values of a given model. The $\fcc$ parameter can provide qualitative descriptions of which elements have more or less prompt/delayed enrichment, but the exact values are uncertain.

\subsection{Increasing the Number of Processes} \label{subsec:k=4}

In our fiducial model, we adopt $K=2$ with the two processes representing prompt, CCSN-like enrichment and delayed, SNIa-like enrichment. While a $K=2$ model can well describe the stellar abundances (e.g., Figures~\ref{fig:all_param1} and~\ref{fig:all_param2}), the abundance residuals cannot be explained by observational noise alone and hold information about the intrinsic variations from a $K=2$ model (\citealp{ness2019} G22, W22, \citealp{ting2022, ratcliffe2023}). Potential sources of such scatter include metallicity-dependent SN yields with a bursty star formation history, environmental variations in the IMF, stochastic sampling of the IMF, and more than two distinct processes (e.g., AGB stars, merging neutron stars, and unique classes of SNIa) with different time delays for enrichment \citep[e.g.][]{belokurov2022, griffith2023}. Note that the existence of many enrichment channels is not in itself sufficient for producing scatter around a $K=2$ model (or even a $K=1$ model); one needs star-to-star variation in the relative amplitude of these channels. For example, in a fully mixed one-zone model, all abundances depend only on time, even if many enrichment channels contribute.

In this Section, we will explore the impact of adding additional processes to our model, increasing from $K=2$ to $K=4$. Because \name{} is sensitive to enrichment with different time delays, adding components could be interpreted as adding sources with distinct enrichment time scales. For example, if AGB stars and SNIa enrich with the \textit{same} time delay, the model would fit both sources in one delayed component. If AGB and SNIa enrich with \textit{different} delay times, a third component could pick up delayed AGB enrichment not captured by the original delayed process. Indeed, evidence of a distinct AGB-like process is identified in G22 and W22, where correlated residuals are used to expand the two-process model. However, both works add components in a restrictive manner that requires choosing which elements to assign to 3rd and/or 4th processes and that does not allow for the original two processes to vary. 

Our goal is to demonstrate the potential for using \name{} to flexibly model more than two enrichment channels and improve the accuracy of the abundance predictions. We allow the model to identify the elements best-fit with additional components and modify the $K=2$ process parameters. Ultimately, such a method could be used to identify elements with more than two enrichment channels, but our data set may not be capable of doing so robustly. In the $K=4$ case, our model becomes
\begin{equation}\label{eq:mij_4}
    m_{ij} = \log_{10}(\Acc\,\qcc + \AIa\,\qIa + A_{i}^{3}\,q_{3, j}^{Z} + A_{i}^{4}\,q_{4, j}^{Z}),
\end{equation}
where $q_{3, j}^{Z}$ and $q_{4, j}^{Z}$ are the third and fourth process vector components and $ A_{i}^3$ and $A_{i}^4$ are the third and fourth process amplitudes. The model, however, does require some regularization to converge. As in the $K=2$ case where we assume that Mg is a pure CCSN element and fix the $\qccFe$ and $\qIaFe$ values, we need elements to regulate our 3rd and 4th processes. We choose Ce and Mn, two elements with larger residuals that likely have additional nucleosynthetic sources---Ce from AGB stars and Mn from distinct classes of SNIa \citep[e.g.][]{gallino1998, reyes2020, gronow2021}. To test the impact of our choice of representative elements we also fit the $K=4$ model with the third and fourth processes fixed to C+N and Cr. We find that similar groups of elements are better fit with additional components. We initialize the $K=4$ model at the $K=2$ model values of $\qcc$, $\qIa$, $\Acc$, and $\AIa$ with the added constraints that 
\begin{equation}\label{eq:q3_z}
    q_{{\rm 3, Mg}}^{\,Z} = 0, \quad 
    q_{{\rm 3, Fe}}^{\,Z} = 0,  \quad 
    q_{{\rm 3, Mn}}^{\,Z} = 0, \quad 
    q_{{\rm 3, Ce}}^{\,Z} = 1 \quad 
\end{equation}
and 
\begin{equation}\label{eq:q4_z}
    q_{{\rm 4, Mg}}^{\,Z} = 0, \quad 
    q_{{\rm 4, Fe}}^{\,Z} = 0,  \quad 
    q_{{\rm 4, Mn}}^{\,Z} = 1, \quad 
    q_{{\rm 4, Ce}}^{\,Z} = 0 \quad 
\end{equation}
at all metallicities. We first fit the $A$-step to only Mg, Fe, Mn, and Ce, and then conduct 32 iterations of the $q$-step and $A$-step, as described in Section~\ref{sec:model}. We again inflate the scatter according to Equation~\ref{eq:inflate_ivar} with $Q=5$. 

\begin{figure*}[htb!]
    \centering
    \includegraphics[width=\textwidth]{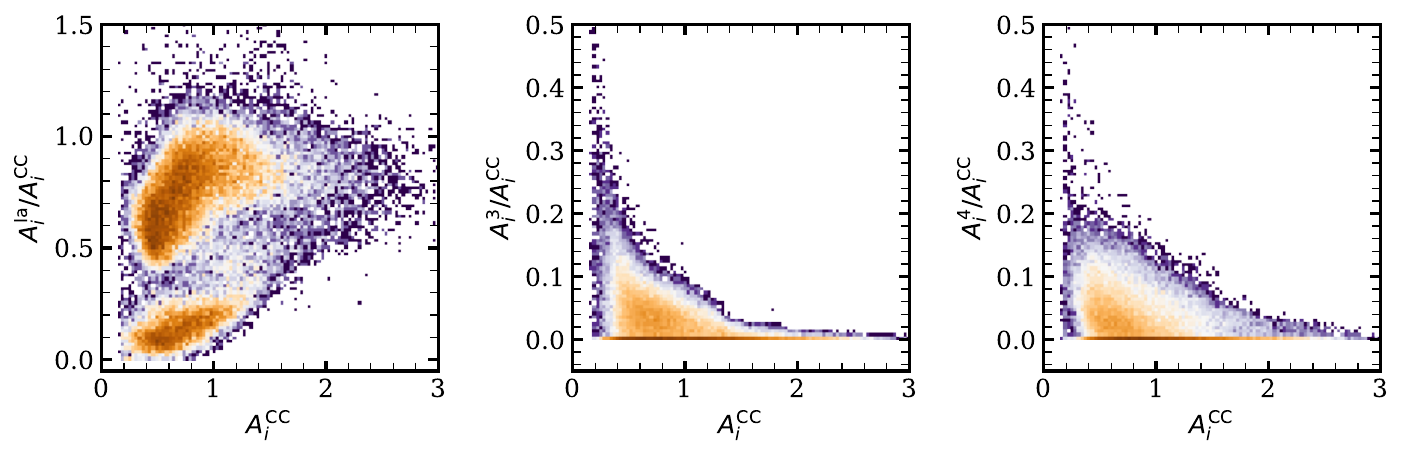}
    \caption{$A^k_i/\Acc$ vs. $\Acc$ distribution for $\AIa$ (left), $A^3_i$ (center), and $A^4_i$ (right). All density plots are logarithmically scaled. The third and fourth processes contribute most to low metallicity stars.}
    \label{fig:A4}
\end{figure*}

\begin{figure*}[htb!]
    \centering
    \includegraphics[width=\textwidth]{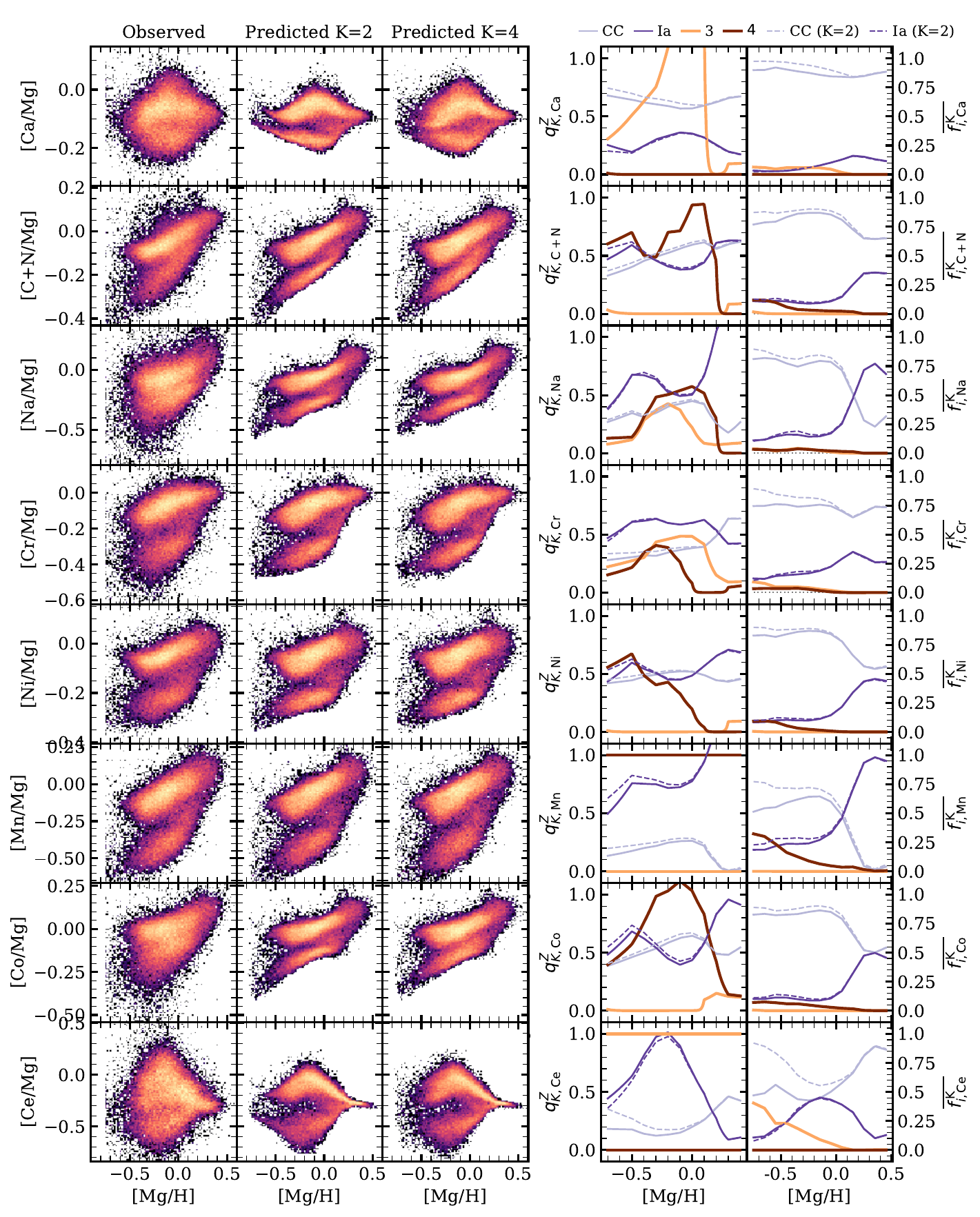}
    \caption{Left: [X/Mg] vs. [Mg/H] abundance distributions for the observed sample (first column), $K=2$ model (second column) and $K=4$ model (third column) for Ca, C+N, Na, Cr, Ni, Mn, Co, and Ce. Observational errors are \textit{not} added to the model predictions. All density plots are logarithmically scaled. Right: process vector components (fourth column) and median low-Ia fractional contribution from each process (fifth column) for the $K=4$ (solid lines) and $K=2$ (dashed lines) models to the low-Ia population as a function of $\mgh$. We plot the median $\qcc$ and $\fcc$ in light purple, $\qIa$ and $f^{\rm Ia}_{ij}$ in dark purple, $q^Z_{3,i}$ and $f^{3}_{ij}$ in light orange, and $q^Z_{4,i}$ and $f^{4}_{ij}$ in dark orange. In the final column we plot a dotted grey line at 0 for reference.} 
    \label{fig:all_paramK}
\end{figure*}

The model converges upon a set of process vector components and amplitudes that can be combined with Equation~\ref{eq:mij_4} to predict the stellar abundances and calculate the fractional contribution from each process. In Figure~\ref{fig:A4}, we plot $A^k_i/\Acc$ vs. $\Acc$ for the SNIa, third, and fourth processes. We find that processes 3 and 4 are most prominent in stars at low metallicity and that there is a large population of stars with $A^3_i$ and/or $A^4_i \approx 0$. In Figure~\ref{fig:all_paramK} we plot the observed and predicted abundance distributions as well as the process vector components and fractional contribution from each component for a subset of elements. We note that the model parameters $q_{3,j}^{Z}$ and $q_{4,j}^{Z}$ should be interpreted in conjunction with the amplitudes, as we set the third and fourth process vector components for Ce and Mn to an arbitrary value with no metallicity dependence. 

We find that the third process, regularized to Ce, contributes at a low level to O, Si, S, Al, and K and more significantly to Ca, Na, Cr, and Ce. The fourth process, regularized to Mn, contributes at a low level to K and more significantly to S, C+N, Na, Cr, Ni, Mn, and Co. These best-fit element groupings resemble, but are not identical to, the elements selected for additional components in W22, where the third process included Ca, Na, Al, K, Cr, and Ce and the fourth process included Ni, V, Mn, and Co. 

In the left-most columns of Figure~\ref{fig:all_paramK}, we plot the $\xmg$ v.s. $\mgh$ distributions for the observed stellar sample, the $K=2$ model predictions, and the $K=4$ model predictions for Ca, C+N, Na, Cr, Ni, Mn, Co, and Ce. We note that the predicted abundances do \textit{not} have noise added (unlike Figures~\ref{fig:all_param1} and~\ref{fig:all_param2}) to highlight the differences between the $K=2$ and $K=4$ predictions. Comparing the predicted abundances from the $K=2$ and $K=4$ process models, we see that the $K=4$ process model is better able to capture the abundance scatter than the $K=2$ model, especially at the low metallicity end of the low-Ia population. This result is expected, as adding more model components will increase the abundance space that \name{} is able to reproduce.

In the fourth and fifth columns of Figure~\ref{fig:all_paramK}, we plot the process vector components and median $f^{k}_{ij}$ as a function of $\mgh$ of the low-Ia population (Equation~\ref{eq:lowIa}), respectively, where
\begin{equation}\label{eq:fX}
    f^{k}_{ij} = \frac{A_{i}^{k} \, q_{k,j}^Z}{\Acc\,\qcc + \AIa\,\qIa + A_{i}^{3}\,q_{3, j}^{Z} + A_{i}^{4}\,q_{4, j}^{Z}}.
\end{equation}
We include $\qcc$ and $\qIa$ as well as median $\fcc$ and $f^{\rm Ia}_{ij}$ from the $K=2$ model in respective columns for comparison. We see that the third process contributes significantly to Ca, Na, Cr, and Ce at low metallicity, with decreasing contribution up to $\mgh\approx0.1$. The fractional contribution from the $K=2$ prompt and delayed processes to these elements decreases under the $K=4$ model. The fourth process behaves in a similar manner but with elements C+N, Na, Cr, Ni, Mn, and Co. The fractional contribution from the third and fourth process is nearly identical in the high-Ia population.

\begin{figure*}[htb!]
    \centering
    \includegraphics[width=\textwidth]{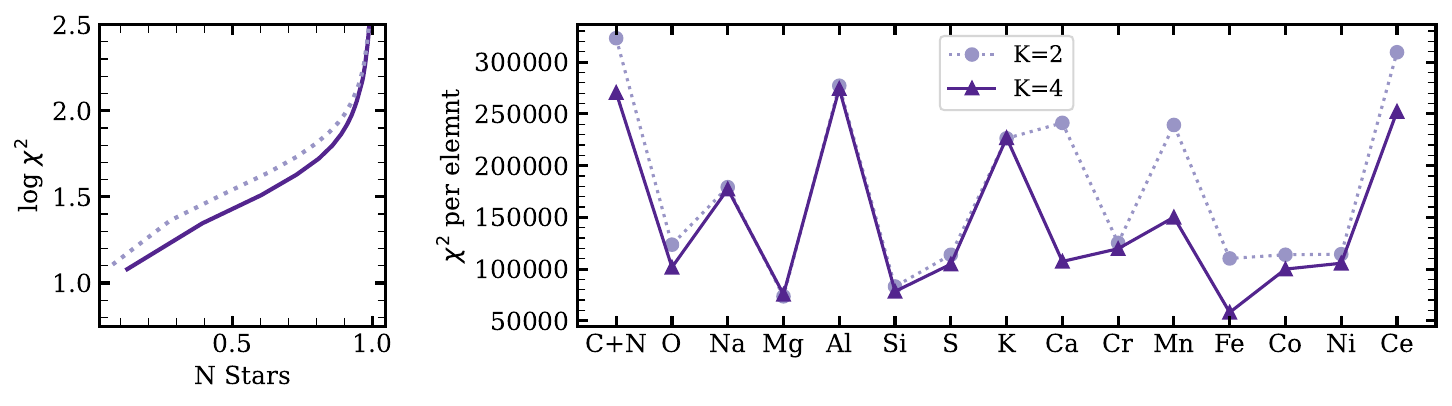}
    \caption{Left: cumulative distribution of $\log_{10} (\chi^2)$ for the $K=2$ model (dotted light purple line) and the $K=4$ model (solid dark purple line). Right: $\chi^2$ per element for the same model fits. Elements are ordered by atomic number.}
    \label{fig:comp_Ks}
\end{figure*}

=The statistical improvement in the \name{} between the $K=2$ and $K=4$ models is evident in the $\chi^2$ values.
In Figure~\ref{fig:comp_Ks}, we plot the cumulative $\log_{10}(\chi^2)$ distributions for the fits to each star and the total $\chi^2$ for each element for the fiducial $K=2$ model and the $K=4$ model. 
We find that the cumulative $\log_{10}(\chi^2)$ distribution decreases with the increase in model components by greater than two for most stars, expected for the addition of two degrees of freedom. 
We also find that the $\chi^2$ per element is lower for all elements in the $K=4$ model. 
Significant improvements to Ca, C+N, Mn, and Ce are likely due to the additional third and fourth components capturing abundance scatter that the original two processes could not. 
Notably, we also see a significant improvement in the Fe fit, even though we require $q_{\rm 3,Fe}^{Z} = q_{\rm 4,Fe}^{Z} = 0$. Because all elements influence the $K=2$ model fit, the fiducial model was likely pulled away from the best solution for Fe to accommodate another element, like Mn. 
With the additional components able to account for the non-Fe-like enrichment, the original two processes are better able to capture the Fe enrichment.

Through this investigation, we find that \name{} is extendable to $K>2$ processes. The additional processes improve the model quantitatively, but additional work is needed to improve the nucleosynthetic interpretability. We provide a discussion of the future science that \name{} and the $K=4$ model enable below.

\section{Discussion}\label{sec:discussion}

In this paper, we present \name{}, a flexible and data-driven model for inferring nucleosynthesis yields. \name{} describes stellar abundances as the sum of $K$ components, where each component is the product of a metallicity-dependent process vector component (fit to each element) and a process amplitude (fit to each star).    Combined with a likelihood function and a set of assumptions (Section~\ref{sec:model}) that make the processes interpretable in terms of nucleosynthetic sources, the best-fit \name{} parameters can be used to calculate fractional contributions from each process as well as a full suite of predicted $K$ process abundances. 

We fit \name{} with $K=2$ to abundances labels for 15 elements and 48,659 RGB stars in APOGEE DR17, selecting a population that minimizes statistical and systematic errors while spanning an [Mg/H] of -0.8 to 0.5.
In the $K=2$ model, the first process, fixed to Mg, represents prompt CCSN-like enrichment, and the second process, fixed to Fe, represents delayed SNIa-like enrichment---but other nucleosynthetic sources with similar time delays may be mixed into each. Under our adopted assumptions, the prompt process also contributes to Fe, but the delayed process does not contribute to Mg, in accordance with theoretical expectations for CCSN and SNIa. Overall, we find that $K=2$ is a good fit to the data and that the model successfully recovers the global abundance patterns in the Milky Way. While \name{} does not rely on $\femg$ vs. $\mgh$ bimodality or median abundance trends, it is able to recover the observed bimodal abundance distribution. Further, the fit parameters $\Acc$ and $\AIa$ act as combined individual-process abundance labels, revealing a clearer signature of bimodality at high metallicity in $\AIa/\Acc$ vs. $\mgh$ space than in $\femg$ vs. $\mgh$. This suggests that the \name{} fit parameters and predicted abundances could be used as a higher signal-to-noise tracer of nucleosynthesis, as they are using a justified likelihood to condense information from 15 elements into two variables.

To test the assumptions of the fiducial model, we explore the impact of varying the fixed value of $\qccFe$.
We find that high values of $\qccFe$ (0.5 and 0.45) are not able to reproduce the observed [Fe/Mg] vs [Mg/H] abundance distribution, regardless of the process vector component's metallicity dependence.
Our requirement that $\Acc$ and $\AIa$ are non-negative makes it impossible for the model to reproduce the lowest $\femg$ values in the APOGEE data. Values $\qccFe = 0.4$ and 0.35 with $\dqccFe = 0$ or 0.15 produces similarly successful fits. While predicted abundance distributions appear similar for these models, the implied fractional contribution from the prompt process is dependent upon the Fe assumption for elements with substantial delayed enrichment. Through this exploration, we conclude that the quantitative nucleosynthetic interpretation of \name{} is dependent upon the input assumptions, and that there is inherent uncertainty in the $\fcc$ values. 

Finally, we expand \name{} from $K=2$ to $K=4$, regularizing the third and fourth processes to Ce and Mn. \name{} builds off of the original model, such that the $K=4$ model starts at the $K=2$ solution and then finds the best-fit parameters for $K=4$, altering the original solution and allowing all elements (except Mg and Fe) to have contributions from additional processes. We find that S, Ca, C+N, Na, Cr, Mn, Co, Ni, and Ce are best fit with a third and/or fourth component, with such processes contributing most significantly at low metallicity. 
The information for constraining the $q^{Z}_{k,j}$ values for the third (fourth) process come from the star-by-star deviations of Ce (Mn) from the $K=2$ model predictions and their correlation with deviations for other elements $\text{X}_i$. Relative to the approach taken in W22 (Section 8), our $K=4$ model requires non-negative $q^Z_{k,j}$ and $A_i^k$ for all elements and stars, and it starts by tying the third and fourth processes to individual elements rather than groups of elements.
The $K=4$ model improves the ability of \name{} to fit the abundances of all elements but especially improves predictions of Ca, C+N, Fe, Mn, and Ce. This successful implementation of a $K=4$ model shows that \name{} can be extended to $K>2$, and it has potential future use in constraining enrichment beyond a single prompt and delayed process---critical to understanding enrichment from AGB stars, merging neutron stars, and rarer novae. 

\name{} is based upon the two-process model developed in W19 and W22. While the two models are identical in format for the $K=2$ case, the model assumptions, parameter derivations, and implementations differ. The W22 two-process model derives process vector components from median abundance trends, reliant upon $\mgfe$ vs. $\mgh$ bimodality, and fits process amplitudes to a subset of $2-6$ $\alpha$ and Fe-peak elements. \name{}, on the other hand, employs a likelihood function fit to all stars and all elements to derive both process amplitudes and vector components. Our more data-driven implementation results in the improved ability of the $K=2$ model fit to predict all of a star's abundances. Notably, \name{} can better predict C+N and Mn abundances than the W22 two-process model, since all elements are used to constrain the fits. The most significant improvement to the original two-process model, though, is in \name{}'s flexibility. The flexible implementation of the model allows us to easily vary the assumptions, such as $\qccFe$, and increase the number of model components to study the impact of our assumptions on the results and push the interpretation of \name{} beyond standard CCSN and SNIa nucleosynthesis in a less restrictive manner than W22 and G22. 

However, \name{} is not without its own faults. The assumptions listed in Section~\ref{sec:model} may incorrectly skew our results, and the model could benefit from improvements in implementation. While assumptions (4) and (5) on Mg and Fe production are flexible, \name{} requires that both elements have fixed process vector components. If our assumptions are incorrect and, for instance, Mg is not a pure prompt element or (in the $K=4$ case) Fe has contributions from multiple delayed sources, our nucleosynthetic interpretation of \name{} may be wrong. This becomes more challenging as $K$ increases and we have to make more assumptions to break rotational symmetries (the symmetries in which the process amplitudes and the process vectors are transformed in corresponding ways to leave the predictions unchanged). Additionally, assumption (7) states that the APOGEE data products can be used for this project, but we inflate outlying-star abundance errors with a softening parameter, $Q$ to account for their likely underestimation. It is also possible that $Q$ is accounting for some of the real intrinsic scatter in the data and inflating the observational error on true outlier stars. In future \name{} implementation, the development of a more robust method to justifiably down-weight outlier stars from the global fits would be beneficial. This method should account for both non-Gaussian observational errors (e.g., from bad telluric subtraction or unlucky line blends) and physically interesting outliers (e.g., from binary mass transfer).
Finally, \name{} fits process vector components along a spline with 11 knots (assumption 6), and those knots have fixed locations in metallicity.
As this method fits a polynomial between each knot, it can result in sharp features at the knot locations in metallicity regions with few points or large scatter. Fitting process vector components with a differentiable function might be more reasonable, though results shown in Appendix C.1 in G22 suggests that this change might have minimal impact on the results. 
In general this model for the metallicity-dependence of the yields is very rigid; a better model could both have more flexibility and be smoother.

Beyond improvements to the underlying model assumptions and implementation, \name{} needs to include parameter uncertainty. While the model delivers process vector components and amplitudes, which can be used to calculate $f^{k}_{ij}$ and $K$ process predicted abundances, the current implementation does not return errors on any variable. The best method to derive such errors has not been explored, but one could use the likelihood function or bootstrapping. These methods will encapsulate the uncertainty on process parameters from the APOGEE abundance errors but will not capture the uncertainty due to model assumptions, such as $\qccFe$ (Section~\ref{subsec:qccFe}). 

While such future changes will improve the model, the current form of \name{} and its data products can support ongoing research and will enable new science.
Most immediately, \name{} provides high signal-to-noise abundance labels, $\Acc$ and $\AIa$, as well as de-noised stellar abundances ($m_{ij}$).
The best-fit values of $\Acc$ and $\AIa$, in particular, are powerful tracers of nucleosynthesis.
They show a bimodality at all metallicities, as do some of the de-noised abundances.
And---because the model is a maximum-likelihood model---they represent information-theory optimal combined measures of $\alpha$ and Fe-peak abundances.
That is, these data-driven amplitudes could replace more theory-driven measures of the relative contributions of CCSN and SNIa enrichment channels.

\begin{figure*}[htb!]
    \centering
    \includegraphics[width=.9\textwidth]{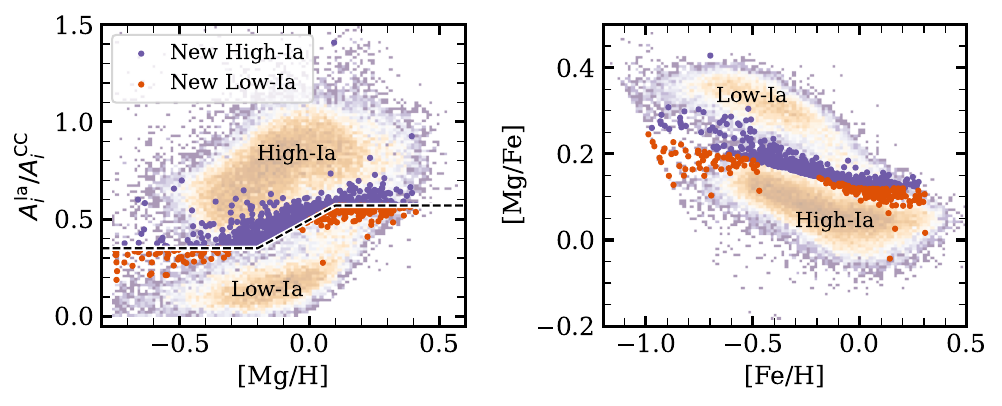}
    \caption{Left: distribution of stars in $\AIa/\Acc$ vs. [Mg/H] where the black dashed line is the dividing line between the high-Ia and low-Ia populations (Equation~\ref{eq:lowIa new}). Stars that are re-classified as high-Ia are shown in purple (647 stars) and stars that are re-classified as low-Ia are shown in orange (224 stars). Right: Same as left, but in [Mg/Fe] vs. [Fe/H] space. The high-Ia and low-Ia populations have been labeled in both panels for clarity. The symbols emphasize the stars at the edge of the populations but the re-classified stars make up only $\sim 2\%$ of the total population.}
    \label{fig:pop_divis}
\end{figure*}

In Section~\ref{subsec:parameters}, we showed that the high-Ia and low-Ia populations are more clearly defined in $\AIa/\Acc$ vs. [Mg/H] than in [Fe/Mg] vs/ [Mg/H]. In amplitude space, the low-Ia population can be re-defined as 
\begin{equation}\label{eq:lowIa new}
\begin{cases}
\AIa/\Acc < 0.35 ,    & \mgh<-0.2 \cr
\AIa/\Acc < 0.5 + 0.7 \, \mgh,  & -0.2\geq\mgh<0.1, \cr
\AIa/\Acc < 0.57 ,    & \mgh\geq0.1 \,. \cr
\end{cases}
\end{equation}
Compared to Equation~\ref{eq:lowIa} (W19, W22), this new definition re-classifies 647 stars as high-Ia and 224 stars as low-Ia. We show the location of these stars in $\AIa/\Acc$ vs. [Mg/H] and [Mg/Fe] vs. [Fe/H] in Figure~\ref{fig:pop_divis}. Many of the re-classified stars are at $\feh>-0.1$. When dividing in [Mg/Fe], it is difficult to correctly separate the populations at high metallicity, as they are blended together. Our new definition also re-classifies many stars near [Fe/H] of -0.3 as high-Ia, suggesting that the W19 and W22 high-Ia definition has too shallow a slope. While only $\sim 2\%$ of stars are re-classified under the new definition, we suggest that Equation~\ref{eq:lowIa new} be used to chemically define the low-Ia and high-Ia populations if \name{} fits are available, especially if studying stars with $\feh > -0.1$. 
Beyond improving the definition of high-Ia and low-Ia populations, the \name{} parameters and predicted abundances could be used in any current analysis that strives to show trends with abundance labels. We predict that trends of stellar parameters with [X/H] will be clearer when comparing to $m_{ij}$, $\Acc$, or $\AIa$.

Such analysis with \name{} parameters will be useful in studying nucleosynthesis, dynamics, disk formation, stellar ages, and much more. However, in this paper we only present fits for a small population with restricted stellar parameters, relative to the full APOGEE sample. 
While \name{} could be fit to the full APOGEE sample, systematic abundance effects with $\teff$ and $\logg$, as well as other abundance artifacts \citep[e.g.,][]{jonsson2020, griffith2021a}, cause the abundance trends to differ across the Hertzsprung-Russell diagram. The best-fit \name{} parameters for the giants would differ from those for the dwarfs. If such systematics could be accounted for (see Sit et al. in prep) we could fit the full APOGEE stellar sample with \name{}, or train \name{} on a subset of high signal-to-noise stars and apply the fits to the full sample. This potential future analysis could reveal additional information about the nucleosynthetic history of our Galaxy and would provide higher signal-to-noise abundance labels for the full sample.

\begin{figure*}[htb!]
    \centering
    \includegraphics[width=.6\textwidth]{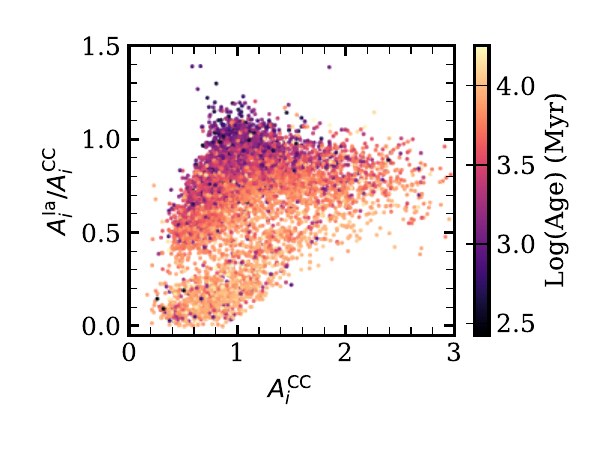}
    \caption{$\AIa/\Acc$ vs. $\Acc$ distribution for stars in the APOKASC sample \citep{pinsonneault2018} Each point is colored by the star's asteroseismic $\log_{10}$(age), with younger stars in black and older stars in yellow. A clear gradient with $\log_{10}$(age) and process parameters exists with young outliers scattered throughout. This shows that the process amplitudes constitute a good-quality age indicator.}
    \label{fig:age}
\end{figure*}

The success of the two-process model (W19, W22) and \name{} with $K=2$ suggests that the distribution of disk stars in APGOEE abundance space is largely two-dimensional (2D), though more dimensions are required to fully explain the data (\citealp{ting2022}, W22). In this paper, we have focused on a 2D nucleosynthetic model, with the two dimensions representing prompt CCSN-like enrichment and delayed SNIa-like enrichment. However, another 2D class of theoretical models for the Milky Way exists, describing stars in terms of birth radius and birth date \citep[e.g.,][]{frankel2018, ness2022}. Are these two 2D models related? If they are, then the nucleosynthetic parameters from \name{} ($\AIa$ and $\Acc$) should predict asteroseismic ages (or masses), up to unpredictable aspects of mass transfer, as well as the guiding center radius, up to unpredictable aspects of radial migration. While a deeper study of the implications of the disk's two-dimensionality is outside the scope of this work, we show the relationship between asteroseismic age from the APOKASC sample \citep{pinsonneault2018} and process amplitudes in Figure~\ref{fig:age}. Here we see a clear gradient in age with $\Acc$ and $\AIa/\Acc$ (as in G22 and W22), though outlier stars are scattered throughout. We predict that the \name{} parameters will be better age diagnostics than APOGEE abundances, and that age outliers may be mass transfer objects. 

Finally, because of the flexibility of \name{}, new scientific applications are enabled that were not feasible before. Because \name{} performs well with a low number of stars and does not rely on [Mg/Fe] vs. [Fe/H] bimodality, non-bimodal populations can now be fit with a multi-component nucleosynthetic model. \name{} could be applied to the low metallicity disk, halo, Gaia Enceladus Sausage, Nubecula Major, Nubecula Minor\footnote{Historically referred to as the LMC and SMC.}, other Milky Way satellites, and more. \name{} can also be easily extended to $K>2$ in a much less restricted way than the two-process model. While a $K=2$ model well describes the global abundance patterns, intrinsic residual scatter on the scale of 0.01 to 0.02 dex remains (\citealp{ting2022}, W22, G22). This scatter could be signatures of enrichment from non-CCSN/SNIa sources, stochastic sampling of the IMF, environmental IMF variations, or metallicity-dependent SN yields with a bursty star formation history \citep[e.g.][]{belokurov2022, griffith2023}. While it is difficult to identify non-CCSN or SNIa enrichment in the APOGEE data alone, where only C+N and Ce are expected to have significant contributions from other sources, there may be signatures in other surveys with better coverage of heavier elements. Applying a $K>2$ model to GALAH \citep{buder2021}, or an overlapping sample of APOGEE and GALAH stars \citep{nandakumar2022}, could prove more successful. 

In the $K=2$ and $K>2$ cases, results from \name{} will help us disentangle our Galactic formation and enrichment history. This data-driven model opens doors to many new research projects and exciting future scientific results. To use \name{} yourself, please reference the \name{} GitHub repository\footnote{\url{https://github.com/13emilygriffith/KProcessModel}} or contact the corresponding author.

\section{Acknowledgements}
It is a pleasure to thank
  Polly Frazer (NYU),
  Adrian Price-Whelan (Flatiron),
  Tawny Sit (OSU),
  Soledad Villar (JHU),
  the Darling research group at CU Boulder,
  CU Boulder Research Computing services and staff,
  and the Astronomical Data group at the Flatiron Institute
for valuable discussions and help.

E.J.G. is supported by an NSF Astronomy and Astrophysics Postdoctoral Fellowship under award AST-2202135.
B.R. acknowledges support by the Deutsche Forschungsgemeinschaft under the 
grant MI 2009/2-1.

Funding for the Sloan Digital Sky Survey V has been provided by the Alfred P. Sloan Foundation, the Heising-Simons Foundation, the National Science Foundation, and the Participating Institutions. SDSS acknowledges support and resources from the Center for High-Performance Computing at the University of Utah. The SDSS web site is \url{www.sdss.org}.

SDSS is managed by the Astrophysical Research Consortium for the Participating Institutions of the SDSS Collaboration, including the Carnegie Institution for Science, Chilean National Time Allocation Committee (CNTAC) ratified researchers, the Gotham Participation Group, Harvard University, Heidelberg University, The Johns Hopkins University, L’Ecole polytechnique f\'{e}d\'{e}rale de Lausanne (EPFL), Leibniz-Institut f{\"u}r Astrophysik Potsdam (AIP), Max-Planck-Institut f{\"u}r Astronomie (MPIA Heidelberg), Max-Planck-Institut f{\"u}r Extraterrestrische Physik (MPE), Nanjing University, National Astronomical Observatories of China (NAOC), New Mexico State University, The Ohio State University, Pennsylvania State University, Smithsonian Astrophysical Observatory, Space Telescope Science Institute (STScI), the Stellar Astrophysics Participation Group, Universidad Nacional Aut\'{o}noma de M\'{e}xico, University of Arizona, University of Colorado Boulder, University of Illinois at Urbana-Champaign, University of Toronto, University of Utah, University of Virginia, Yale University, and Yunnan University.

\software{matplotlib \citep{hunter2007}, NumPy \citep{harris2020}, pandas \citep{pandasa, pandasb}, astropy \citep{astropy2013, astropy2018, astropy2022}, and jax \citep{jax}.}

\facilities{Sloan, Kepler}

\bibliography{kpm}{}
\bibliographystyle{aasjournal}

\end{document}